\documentclass[a4paper,11pt]{article}

\usepackage[margin=1in]{geometry} 

\usepackage{amsmath}
\usepackage{amsthm}
\usepackage{amssymb}
\usepackage{lineno}
\usepackage{subcaption}  

\usepackage{placeins}

\usepackage{etoolbox}

\usepackage{graphicx}
\usepackage{bm}
\usepackage{lineno}
\usepackage{caption}

\usepackage{booktabs}

\usepackage[breaklinks]{hyperref}

\usepackage{multirow}

\usepackage{comment}
\usepackage{xargs}                      

\newcommand\R{\mbox{\text{Re}}}

\usepackage[utf8]{inputenc}

\hypersetup{
	unicode,
	colorlinks=true,
	linkcolor=blue,
	filecolor=magenta,      
	citecolor=blue,
	urlcolor=cyan,
	pdfauthor={ M. A. Mendez},
	pdftitle={Linear and Nonlinear Dimensionality Reduction from Fluid Mechanics to Machine Learning},
	pdfsubject={Linear and Nonlinear Dimensionality Reduction from Fluid Mechanics to Machine Learning},
	pdfkeywords={article, template, simple},
	pdfproducer={LaTeX},
	pdfcreator={pdflatex}
	colorlinks=true,
	citecolor=blue,
	linkcolor=blue,
	filecolor=magenta,      
	urlcolor=blue,
}

\usepackage[most]{tcolorbox}
\usepackage{booktabs}
\usepackage{amsmath}
\usepackage{wrapfig}

\usepackage{mathtools}
\tcbset{enhanced,colback=cyan!2!white,colframe=cyan!90!white,coltitle=blue!20!black,fonttitle=\bfseries}

\usepackage{listings}
\usepackage{hyperref}
\hypersetup{
	colorlinks=true,
	linkcolor=blue,
	filecolor=magenta,      
	citecolor=blue,
	urlcolor=cyan,
}

\usepackage[sort&compress,authoryear,round]{natbib}

\usepackage{algorithm, algpseudocode} 
\usepackage{mathrsfs} 

\title{\LARGE \textbf{Nonlinear System Identification for Model-Based Control of Waked Wind Turbines}}

\author{
  Sebastiano Randino$^{1,2}$\thanks{Corresponding author: sebastiano.randino@vki.ac.be} \and
  Lorenzo Schena$^{1,5}$ \and
  Nicolas Coudou$^{1}$ \and
  Emanuele Garone$^{2}$ \and
  Miguel Alfonso Mendez$^{1,3,4}$
}

\date{%
\begingroup
\small
\textit{%
$^1$ Environmental and Applied Fluid Dynamics, von Karman Institute for Fluid Dynamics, Belgium \\
$^2$ SAAS, Université Libre de Bruxelles, Belgium \\
$^3$ Aero-Thermo-Mechanics Laboratory, École Polytechnique de Bruxelles, Université Libre de Bruxelles, Belgium \\
$^4$ Experimental Aerodynamics and Propulsion Lab, Universidad Carlos III de Madrid, Spain \\
$^5$ Department of Mechanical Engineering, Vrije Universiteit Brussel (VUB), Belgium
}
\endgroup
}

\begin{document}
\maketitle
	
\begin{abstract}

This work presents a nonlinear system identification framework for modeling the power extraction dynamics of wind turbines, including both freestream and waked conditions. The approach models turbine dynamics using data-driven power coefficient maps expressed as combinations of compact radial basis functions and polynomial bases, parameterized in terms of tip-speed ratio and upstream conditions. These surrogate models are embedded in a first-order dynamic system suitable for model-based control. Experimental validation is carried out in two wind tunnel configurations: a low-turbulence tandem setup and a high-turbulence wind farm scenario. In the tandem case, the identified model is integrated into an adapted $K\omega^2$ controller, resulting in improved tip-speed ratio tracking and power stability compared to BEM-based and steady-state models. In the wind farm scenario, the model captures the statistical behavior of the turbines despite unresolved turbulence. The proposed method enables interpretable, adaptive control across a range of operating conditions without relying on black-box learning strategies.

\vspace{7mm}
\end{abstract}
	
\section{Introduction}
\label{sec:1}

Wind energy is the fastest-growing renewable sector and a cornerstone of global decarbonization efforts. In Europe, the installed capacity has doubled from 128 GW in 2014 to 255 GW today \citep{ewea2014statistics,windeurope2025}. To meet rising demand, turbines have grown over tenfold in size and are increasingly clustered into large wind farms \citep{caduff2012wind,porte2020wind}. As wind speed varies, wind turbines are controlled to maximize energy capture at low wind speeds, while constraining power output and mitigating structural loads at high wind speeds \citep{wright2008advanced}.

Control of wind turbines has received enormous attention in recent years, with research increasingly focusing on advanced strategies. Approaches such as Fuzzy Logic Control (FLC,~\citet{Maafa2024,Bharani2022,borunda2024intelligent}), Sliding Mode Control (SMC~\citet{Zribi2017,Echiheb2022}), Maximum Power Point Tracking (MPPT~\citet{Pande2023,Abdullah2012}), and Model Predictive Control (MPC~\citet{schlipf2013nonlinear,GonzalezSilva2022})—have been proposed to enhance robustness and adaptability under varying conditions. However, their adoption remains limited due to practical challenges such as the need for expert tuning (FLC), high-frequency oscillations (SMC), sensitivity to wind fluctuations (MPPT), and computational demands (MPC). In recent years, AI-driven control approaches have emerged as a compelling alternative, leveraging data to design or adapt controllers capable of handling nonlinear, poorly modeled turbine dynamics \citep{tomin2019intelligent,sierra2020exploring}. Methods based on Artificial Neural Networks (ANN,~\citet{majout2024artificial}), Adaptive Neuro-Fuzzy Inference Systems (ANFIS,~\citet{chhipa2021adaptive}), Reinforcement Learning (RL,~\citet{schena2022wind,soler2024reinforcement}), and Evolutionary Algorithms (GA,~\citet{guediri2023modeling}, PSO~\citet{khurshid2022optimal}) have demonstrated promising performance in simulations. Nonetheless, their application in operational turbines remains constrained by stringent industry requirements for reliability, consistency, and certifiability \citep{Chatterjee_2021}. Emerging directions such as Safe Reinforcement Learning \citep{gu2024reviewsafereinforcementlearning} seek to address these barriers, although real-world deployment is still in its early stages.

Despite the growing interest in advanced controllers, classical model-based strategies remain the standard in industrial wind turbines due to their robustness and simplicity \citep{wright2008advanced,pao2009tutorial}. Power maximization is traditionally carried out with torque controllers following the so-called $K\omega^2$ law, which adjust the generator torque while maintaining a fixed, optimal blade pitch, whereas load mitigation is typically assigned to a Proportional-Integral (PI) controller of the blade pitch to sustain rated power output \citep{laks2009control}. These typically rely on first-order nonlinear models, whose accuracy critically impacts performance. For example, the gain $K$ in the $K\omega^2$ controller depends on the optimal tip-speed ratio \citep{pao2009tutorial}, while PI controllers are tuned using linearized turbine dynamics \citep{abbas2021reference}. A persistent challenge is the identification of the rotor’s nonlinear aerodynamic behavior \citep{ribeiro2023nonlinear}. Common approaches include generating aerodynamic maps via momentum theory—most notably the Blade Element Momentum (BEM) method—or using pre-defined functional forms (e.g., polynomials or sinusoids) calibrated from experimental data \citep{pao2009tutorial,castillo2023comparison}. Accurate aerodynamic modeling is critical for minimizing energy losses in wind turbine control, with errors potentially reducing large-scale production by up to 3\% \citep{pao2009tutorial}. Moreover, model accuracy degrades over time due to structural aging, blade erosion, and ice formation \citep{staffell2014does,campobasso2023probabilistic}. The continuous growth of turbine size and increasing wake interactions in large wind farms further challenge models originally designed for isolated turbines in steady inflow \citep{pao2009tutorial,shapiro2021turbulence}: unsteady wake-induced conditions violate key assumptions of methods like BEM \citep{leishman2002challenges}, while blade flexibility alters aerodynamic load distributions beyond what standard models capture \citep{cao2022comparing}. Though dynamic BEM introduces empirical corrections \citep{papi2023going}, persistent discrepancies with real-world behavior continue to motivate research into adaptive control and system identification \citep{pusch2024optimal}.

To address the limitations of first-principles models under real-world conditions, system identification techniques are increasingly employed in wind turbine control \citep{nelles2020nonlinear,bossanyi2012advanced}. These data-driven approaches replace detailed physical modeling with parameter estimation from operational data \citep{rai2017wind}. Most methods adopt linear state-space representations around specific operating points \citep{houtzager2010system}. Examples include linear time-periodic (LTP) models identified via harmonic transfer functions using HAWC2 simulations \citep{allen2011output}, and linear time-invariant (LTI) models estimated through Predictor-Based Subspace Identification (PBSID) \citep{houtzager2010system} or Frequency Domain Decomposition of field data \citep{jasniewicz2011wind}. Simple Auto-Regressive (AR) models have also shown competitive accuracy with minimal parameters \citep{mahmoud2012model}. These linear models frequently support advanced control schemes such as Model Predictive Control (MPC) \citep{verwaal2015predictive}. Other strategies include adaptive second-order transfer functions \citep{simani2013data} and switching between LTI models across operating regions, with fuzzy logic used to smooth transitions \citep{jelavic2006identification}.

The limitations of linear control strategies under switching conditions are discussed in \cite{gros2017real}. Nonlinear models have demonstrated improved performance over traditional Gain-Scheduled PI controllers \citep{kumar2010simulating}, motivating system identification efforts using techniques such as Variable-Order Fractional Neural Networks \citep{aslipour2019identification}, fuzzy systems \citep{simani2012application}, and neural networks \citep{kelouwani2004nonlinear}. Clustering-based piecewise affine models have also been explored for power regulation \citep{sindareh2021machine}. However, concerns over generalization and reliability of these black-box methods have motivated hybrid methods that embed physical insights into data-driven models, improving robustness across operating conditions \citep{gebel2025system}.

A promising direction is the identification of physically interpretable quantities like power or thrust coefficients. Common approaches use polynomial models with Recursive Least Squares (RLS) for real-time updates \citep{monroy2006real, hosseinpour2017improving}, Reduced Multivariate Polynomial models \citep{son2009estimation}, or observer-based methods for dynamic estimation \citep{yap2012online}. Adaptive systems such as ANFIS have also been applied to optimize power coefficient estimation and enhance performance \citep{petkovic2013adaptive, asghar2017estimation}. Most adaptive modeling efforts for estimating the power coefficient have focused on isolated turbines and relied on numerical simulations where the true coefficient is known \citep{monroy2006real, hosseinpour2017improving, petkovic2013adaptive}. To the authors' knowledge, no studies have retrieved experimentally the power coefficient in the presence of turbine wakes, where conventional models typically fall back on simplified, low-fidelity wake descriptions \citep{bhatt2022stochastic}. A notable exception is \cite{annoni2015system}, who identified a two-turbine array offline using linear transfer functions from experimental data. However, real-time assimilation of the power coefficient in multi-turbine setups remains unaddressed.

This study addresses this gap by proposing the real-time experimental assimilation of the power coefficient for a two-turbine system, with one turbine operating within the wake of the other. The approach identifies a nonlinear, first-order dynamic model of the wind turbine array configuration, suitable for model-based control applications. As a demonstrative application, we consider a small scale set-up of two 0.15 m diameter turbines, challenging the model to recover also low Reynolds number effects, and integrate the identified model into an adapted version of the $K\omega^2$ controller.  

This article is organized as follows. First, the system identification problem of a wind turbine is reviewed in Section \ref{sec:2} while section \ref{sec:3} introduces the proposed data driven closure laws with the identification strategy. Section \ref{sec:4} described the experimental set up. Section \ref{sec:5} presents the results while Section \ref{sec:6} closes with conclusions and perspectives.

\begin{figure}[t]
    \centering    \includegraphics[width=0.60\columnwidth]{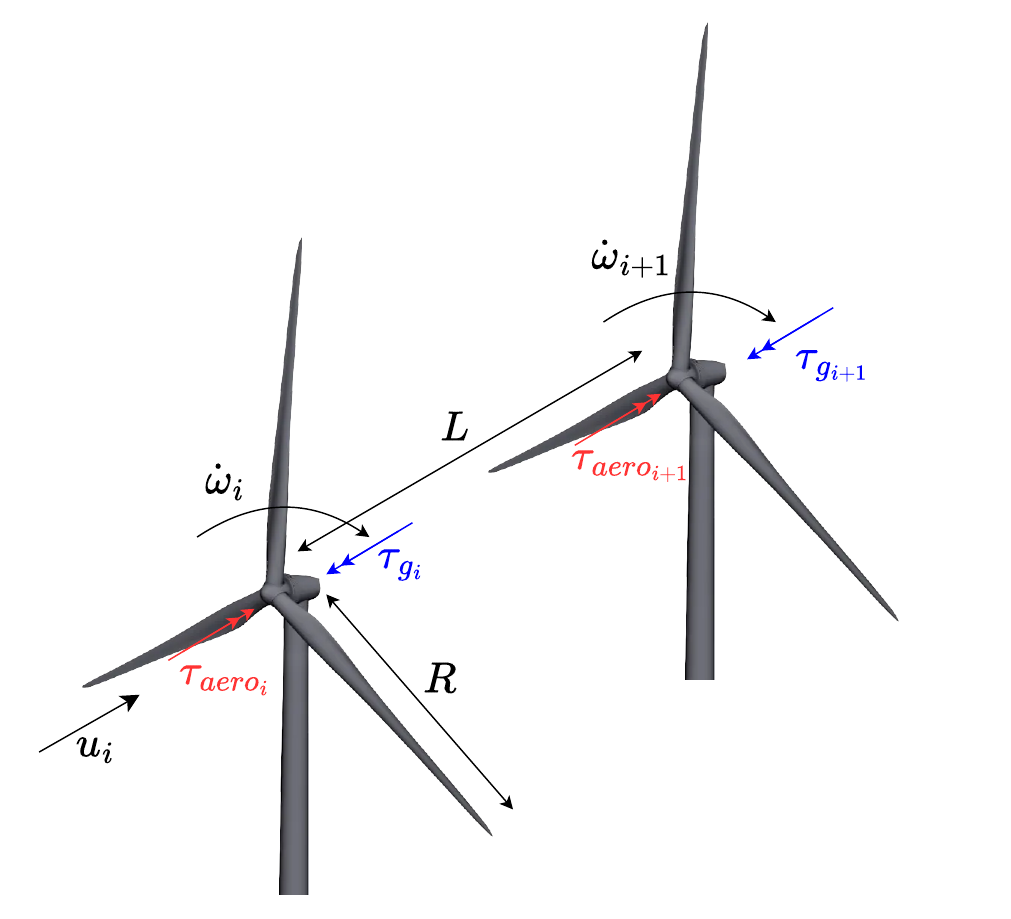}
    \caption{Configuration of interest and relevant variables.}
    \label{fig1}
\end{figure}

\section{System Identification Framework}
\label{sec:2}

The configuration of interest is illustrated in Figure~\ref{fig1}. It consists of identical, aligned turbines with radius \(R\), spaced at a distance \(L\), rotating at angular velocities \(\omega_i\), and subjected to an incoming flow with hub-height equivalent wind speeds \(u_i\). The subscripts \(i\) and \(i+1\) refer to the upstream and downstream waked turbines, respectively, while the subscript \(1\) refers to the first turbine facing the freestream wind speed. We assume that the turbines operate under a power maximization objective in the torque-controlled region; therefore, the only relevant control inputs are the generator torques \(\tau_{g,i}\). 

The aerodynamic torque generated by the turbines is usually related to the power coefficient $C_p$, so that the simplest (nonlinear) ODE model of each turbine can be written as 

\begin{equation}
\label{eq:system_turbine1}
\begin{cases}
\dot{\omega}_1 = f(\omega_1, u_1 , \tau_{g,1} ; C_{p,1}) \\
C_{p,1}  = \tilde{g}_1(\omega_1, u_1 ;\mathbf{w}_1) \\
\omega_1(0) = \omega_{1,0}
\end{cases}
\end{equation} for turbine 1 and 

\begin{equation}
\label{eq:system_turbine2}
\begin{cases}
\dot{\omega}_i = f(\omega_i, u_i(u_1,C_{p,i-1}) , \tau_{g,i} ; C_{pi}) \\
C_{p,i}  = \tilde{g}_i(\omega_i, u_i(u_1,C_{p,i-1}) ;\mathbf{w}_i) \\
\omega_i(0) = \omega_{i,0}
\end{cases}
\end{equation} for a generic waked turbine $i = 2, 3, \dots$, where the parameters $\mathbf{w}_i \in \mathbb{R}^{n_{q}}$ act as closure parameters. The dot denotes time derivatives. Notice that the hub-speed velocity $u_i$ for the waked turbine is linked to the free-stream velocity $u_1$ and the performance of the preceding turbine $i-1$, specifically through its power coefficient $C_{p,i-1}$.

The function $f$ for the generic turbine $i$ reads 

\begin{align}
f(\omega_i, u_1, \tau_{g,i}; C_{p,i}) 
&= \frac{1}{J} (\tau_{a,i} - \tau_{g,i}) \notag \\
&= \frac{1}{J} \left( 
    \frac{1}{2} \rho \pi R^2 \frac{C_{p,i}}{\omega_i} u_1^3 
    - \tau_{g,i} 
\right)
\label{eq:omega_dynamics}
\end{align}  where $J$ is the rotor's moment of inertia [kg·m$^{2}$], $\rho$ is the air density [kg·m$^{-3}$], and the generator is assumed to be directly coupled to the rotor (here, we assume unit efficiency $\eta = 1$ and no gear ratio $N_g = 1$).

The power coefficient of a wind turbine depends nonlinearly on its operating conditions. For a given geometric configuration and blade pitch angle, and at sufficiently high Reynolds numbers \( \R = u_1 D / \nu \), with \( \nu \) denoting the air’s kinematic viscosity, the power coefficient is typically expressed as a function of the tip-speed ratio \( \lambda_i = \omega_i R / u_1 \). In this work, we assume that the power available in the wake depends solely on the amount of power extracted by the preceding turbine. Since the extracted power is linked to the power coefficient of the preceding turbine, which in turn depends only on its tip-speed ratio, i.e., \( C_{p,i-1} = C_{p,i-1}(\lambda_{i-1}) \), this modeling assumption allows us to express the power coefficient of a generic turbine as a function not only of its own tip-speed ratio \( \lambda_i \), but also directly of the tip-speed ratio of the upstream turbine \( \lambda_{i-1} \).

The nonlinear system identification problem addressed in this work consists in modeling the power coefficients \( C_{P,i} \) in \eqref{eq:system_turbine1} and \eqref{eq:system_turbine2} as general parametric functions \( \tilde{g}_i(\omega_i, u_1; \mathbf{w}_i) \). Denoting by \( \mathbf{w} = [\mathbf{w}_1, \mathbf{w}_2, ...] \in \mathbb{R}^{n_q} \) the concatenated parameter vector of the total number of turbines, the joint identification is performed by minimizing the discrepancy between observed data and the predictions of the dynamical model defined in \eqref{eq:system_turbine1} and \eqref{eq:system_turbine2}.

Following a traditional variational approach for nonlinear identification \citep{schena2023reinforcement,Marques2023}, the optimal parameters are those that minimize the integral cost functional:
\begin{equation}
\label{eq:optimization}
\mathcal{J}_{w,i}(\mathbf{\mathbf{w}_i}) = \frac{1}{T}\int_0^{T_0} \mathcal{L}_p\left(\omega_i(t; \mathbf{w}_i), \omega_i^*(t)\right) \, dt\,,
\end{equation} where \( \mathcal{L}_p \) denotes the Lagrangian quantifying the mismatch between the simulated state \( \omega_i(t; \mathbf{w}_i) \) and the measured states \( \omega_i^*(t) \). The time interval \( T_0 \) represents the observation window over which measurements are collected, also referred to as an \emph{episode}, following the terminology in \cite{schena2023reinforcement,Marques2023}. As detailed in the following section, the identification for the different turbines is carried out independently by solving separate minimization problems.

\section{Power Coefficient Closure Laws} \label{sec:3}

The power coefficients $C_{p,i}$ in \eqref{eq:system_turbine1} and \eqref{eq:system_turbine2} are herein modeled as follows 

\begin{equation}
\label{CP1}
C_{p,1}(\lambda_1,\R)=\sum^N_n \sum^M_m w_{1,m,n} \phi_n (\lambda_1;\lambda^*_m,c_m) \psi_m (\R; n)\,,
\end{equation} and 

\begin{equation}
\label{CP2}
C_{p,i}(\lambda_i,\lambda_{i-1})=\sum^N_n \sum^M_m w_{i,m,n} \phi_n (\lambda_i;\lambda_m^*,c_m) \psi_m (\lambda_{i-1};n)\,.
\end{equation} where all the closure coefficients \( w_{i,m,n} \) are stored in the vector \( \mathbf{w}_i \) for each wind turbine, as previously described. The functions \(\phi_n\) and \(\psi_m\) denote radial basis functions (RBFs) and polynomial functions, respectively, as described in detail below. The integers \(M\) and \(N\) represent the total number of basis used.

As demonstrated in the following, the proposed formulation allows for representing a realistic nonlinear power coefficient curve while maintaining a linear dependency on the model parameters, which greatly simplifies the identification procedure. The dependence on the Reynolds number for the first turbine arises from the small scale of the experimental models used (see \citet{bastankhah2017new}), but is known to be negligible for full-scale wind turbines \citep{saint2020parametric}. For the waked turbine, additional Reynolds number effects are assumed to be negligible, as experimental evidence suggests that its behavior is more strongly influenced by the upstream tip-speed ratio \( \lambda_{i-1} \).

The functions $\phi_n(x,\lambda^*_m,c_m)$ and $\psi_m(x,n)$ act as basis functions. The first is a basis of compact support Radial Basis Functions (RBFs)  

\begin{equation}
\phi_m\left(\lambda; \lambda_m^*, c_m\right)= 
\begin{cases}
\left(1-\frac{d^2_m(x)}{c_m^2}\right)^5 & \text{if } |d_m| \leq c_m \\
0 & \text{otherwise}
\end{cases}
\end{equation} where $d_m(x) = \lambda - \lambda_m^*$ is the distance from the RBF center $\lambda_m^*$, and $c_m$ is the basis radius. Five of these bases were collocated in the range of admissible tip-speed ratios at centers $\lambda^*_m=[4,5,6,7,8]$ and having all the same shape factor $c_m=1.5$. The second is polynomial basis up to order $2$ was chosen, hence $\psi_n(x,n)=x^n$ with $n=[0,1,2]$.

\subsection{Initial guess for model parameters}
\label{sec:3.1}

An estimate of the coefficients $w_{i,m,n}$ can be computed using the standard least squares method, under the assumption of ideal steady-state condition $\dot{\omega}=0$. In this condition, from \eqref{eq:omega_dynamics} one has $\tau_{a,i}=\tau_{g,i}$, hence

\begin{equation}
\label{CP}
C_{p,i}=\frac{2 \tau_{g,i} \omega_i }{\rho \pi R^2 u_i^3} \,.
\end{equation}

Assuming that both $\omega_i$ and $u_i$ are measured, equation \eqref{CP} can be used to compute $C_{p,i}$, allowing for deriving initial guesses to train the surrogate models in \eqref{CP1} and \eqref{CP2} via least squares regression. For instance, in the case of the first turbine in \eqref{CP1}, collecting a grid of $n_{\lambda}\times n_{Re}$ training samples, with $n_{\lambda}$ and $n_{Re}$ the number of tip speed ratios and Reynolds numbers in the grid, and enforcing \eqref{CP} in each of the grid points yields a matrix factorization of the form
\begin{equation}
\label{CP_1_Sys}
\bm{C}_{p,1} = \bm{\Phi}(\bm{\lambda}_1) \, \bm{w}_1 \, \bm{\Psi}^T(\textbf{\R}),
\end{equation}
where $\bm{C}_{p,1} \in \mathbb{R}^{n_{\lambda}\times n_{Re}}$ is the matrix of power coefficients computed from \eqref{CP} at all training points. The matrix $\bm{\Phi}(\boldsymbol{\lambda}_1) \in \mathbb{R}^{n_\lambda \times 5}$ is the radial basis function (RBF) matrix evaluated at the tip-speed ratios $\boldsymbol{\lambda}_1$, while $\bm{\Psi}(\textbf{\R}) \in \mathbb{R}^{n_{Re} \times 3}$ contains the polynomial basis evaluated at the Reynolds numbers $\textbf{\R} \in \mathbb{R}^{n_{Re}}$. The coefficient matrix $\bm{w}_{1} \in \mathbb{R}^{5 \times 3}$, with entries $w_{1,m,n}$, contains the parameters to be identified in the surrogate model \eqref{CP1} and de facto encodes the performances of the turbine under the dynamical system in \eqref{eq:system_turbine1}-\eqref{eq:omega_dynamics}.

These coefficients can be computed by pseudo-inverse on both sides of \eqref{CP_1_Sys}:
\begin{equation}
\label{C1_0}
\bm{w}_{1,0} = (\bm{\Phi}^T\bm{\Phi})^{-1}\bm{\Phi}^T \bm{C}_{p,1} \bm{\Psi} (\bm{\Psi}^T\bm{\Psi})^{-1}\,.
\end{equation} where \( \bm{w}_{1,0} \) emphasizes that these are the initial guesses for the optimization.

Considering that this estimate acts only as an initial guess, no regularization was used in the least square solution. On the other hand, taking into account the measurement uncertainties was found to significantly improve the initial guess. These were introduced using a classic weighted formulation on the regression of the RBF block, which reads 

\begin{equation}
\label{C1}
\bm{w}_{1,0} = (\bm{\Phi}^T\mathbf{N}\bm{\Phi})^{-1}\bm{\Phi}^T \mathbf{N} \bm{C}_{p,1} \bm{\Psi} (\bm{\Psi}^T\bm{\Psi})^{-1}\,, 
\end{equation} where the matrix $\mathbf{N}$ is an estimate of the measurement precision, giving more weight to points with lower uncertainty and defined as 

\begin{equation}
\label{Sigma}
\mathbf{N} = \sigma^2_{\min}
\begin{bmatrix}
\frac{1}{\sigma^2_1} & 0 & \dots & 0 \\
0 & \ddots & 0 & \vdots \\
\vdots & 0 & \ddots & 0 \\
0 & \dots & 0 & \frac{1}{\sigma^2_{n_j}}
\end{bmatrix},
\end{equation}
where \( \sigma^2_j \) is the variance of each measurement and \( \sigma_{\min}=\text{min}\{\sigma_i\} \). These were used by propagating the measurement uncertainties of $u_i$, $\tau_{g,i}$ and $\omega_i$ through \eqref{CP} using standard first order approach. The solution of the initial set of coefficients from \eqref{C1} is equivalent to first carry out a weighted least square regression of the model $\mathbf{C_{p,1}}= \bm{\Phi} \mathbf{A}$, followed by a traditional least square regression of the model $\mathbf{A}^T= \bm{\Psi} \bm{w}^T_{1,0}$.

\subsection{A note on stability}\label{sec3p2}

In a physically valid wind turbine model, the loci of steady state solutions $\tau_{a,j}=\tau_{g,j}$ in \eqref{eq:omega_dynamics} must be asymptotically stable, since the net torque decreases with increasing rotor speed in the neighborhood of equilibrium. The stability of the steady state solutions implies that the Jacobian 

\begin{equation}
\label{eq:stability}
\frac{df}{d\lambda_i}
= \frac{1}{J} \left[
\frac{1}{2} \rho \pi R^3 u_i^2 \cdot \frac{d}{d\lambda_i} \left( \frac{C_p(\lambda_i, \mathrm{Re})}{\lambda_i} \right)
- \frac{d\tau_g}{d\lambda_i}
\right]
\end{equation} has to be negative regardless of the control law $\tau_g$ in the steady state conditions. Although this stability condition could be enforced by formulating the identification of $C_p(\lambda,Re)$ in \eqref{CP1} or \eqref{CP2} as a constrained optimization problem, this added complexity proved unnecessary because the structure of the cost function inherently penalizes parameter choices that violate the stability condition. Specifically, parameter choices that violate this stability condition naturally result in turbine models where the desired steady-state solution is no longer the system's attractor. Instead, the system trajectories tend to converge to zero rotational speed, leading to large discrepancies with the data and, consequently, high identification cost. This implicitly discourages such parameter values without the need to explicitly enforce stability constraints during optimization.

\subsection{Adjoint Based Real Time Identification}
\label{sec:3.3}
In this work, the Lagrangian in \eqref{eq:optimization} is defined as
\begin{equation}
\mathcal{L}_{p}(\mathbf{w}_i) = \left( \omega_i^*(t) - \omega_i(t; \mathbf{w}_i) \right)^2\,.
\end{equation}

The optimization for each turbine is performed independently. This is justified by the fact that the downstream turbine model depends solely on the tip-speed ratio of the upstream turbine, \(\lambda_{i-1}\), and is otherwise decoupled from its dynamics. The measured time series \(\omega_i^*(t)\) is filtered using a $4^{\text{th}}$ order Butterworth low-pass filter with a cut-off frequency of \(f_{\text{c.o.}} = 2\,\text{Hz}\) chosen based on the turbine's time response to attenuate high-frequency noise.

The cost functions \( \mathcal{J}_{w,i}(\mathbf{w}_i) \) are optimized using the Adam optimizer \citep{kingma2014adam}, preferred over quasi-Newton alternatives such as the BFGS because of its higher robustness to gradient noise. The gradient \( \nabla_{\mathbf{w},i} \mathcal{J}_{w,i} \) was computed via an adjoint-based formulation \citep{sengupta2014efficient} to avoid solving \( n_w \) additional ODEs per evaluation. The adjoint method computes the gradient as
\begin{equation}
\nabla_{\mathbf{w,i}} \mathcal{J}_{w,i}(\mathbf{w}_i) = \int_0^{T_0} s_\lambda^T (t) \left( \frac{\partial f}{\partial \mathbf{w,i}} \right) \, dt
\end{equation}
where \( s_\lambda(t) \) are the adjoint variables. These are obtained by solving the adjoint terminal value problem 

\begin{equation}
\begin{cases}
\displaystyle \frac{ds_\lambda}{dt} = - \left(\frac{\partial f}{\partial \omega_i}\right)^\top s_\lambda(t) 
                                      - \left( \frac{\partial \mathcal{L}_p}{\partial \omega_i} \right)^\top, \\[8pt]
\displaystyle s_\lambda(T_0) = 0
\end{cases}
\label{eq:adjoint}
\end{equation}
 where all terms are evaluated at the current guess for the states \( \omega_i(t; \mathbf{w}_i) \), which thus requires first a forward integration. We note that the transposition in \eqref{eq:adjoint} is included for consistency with high dimensional problems and is irrelevant to the problem at hand since both Jacobians are scalars. The adjoint-based evaluation requires one forward pass and one backward pass, regardless of the number of unknown parameters $\mathbf{w}_i$. 

Starting from the initialization as in \eqref{C1}, the gradient computation is used then to update the weight guess using the ADAM optimizer \citep{kingma2014adam}

\begin{equation}
    \mathbf{w}_i^{(n+1)} = \mathbf{w}_i^{(n)} - \eta \frac{\hat{\mathbf{m}}^{(n)}}{\sqrt{\hat{\mathbf{v}}^{(n)}} + \epsilon}
\end{equation}
where \( n \) is the iteration number, \( \eta \) is the learning rate, \( \epsilon \) is a small constant to prevent division by zero, and \( \hat{\mathbf{m}} \) and \( \hat{\mathbf{v}} \) are filtered momentum parameters computed as follows
\begin{equation}
\hat{\mathbf{m}}^{(n)} = \frac{\beta_1 \mathbf{m}^{(n-1)} + (1 - \beta_1) \nabla_\mathbf{\mathbf{w,i}} \mathcal{J}_{w,i}^{(n)}}{1 - \beta_1^n}
\end{equation}
\begin{equation}
\label{grad_SMOOTH}
\hat{\mathbf{v}}^{(n)} = \frac{\beta_2 \mathbf{v}^{(n-1)} + (1 - \beta_2) \left[\nabla_\mathbf{\mathbf{w,i}} \mathcal{J}_{w,i}^{(n)}\right]^2}{1 - \beta_2^n}\,,
\end{equation} were \( \mathbf{m} \) and \( \mathbf{v} \) are the first and second moment estimates of the gradient, while \( \beta_1 \) and \( \beta_2 \) are the decay rate coefficients for the moments. 

Multiple restarts were employed to reset the optimization process whenever a sudden jump in the cost function occurred, typically occurring because the current parameter guess violates the stability condition in \eqref{eq:stability}. In such case, the optimization is re-started to re-set the low-pass filtering of the gradient. Additionally, a learning rate schedule was designed to reduce the learning rate once the cost function falls below a predefined threshold. This approach allows the optimizer to take smaller steps and prevent overshooting.

\section{Experimental Methodology}
\label{sec:4}

The detailed experimental setup used to acquire data and interact with the system in real time is described in Section~\ref{sec:expsetup}. The identification methodology is evaluated across multiple scenarios. The first, detailed in Section~\ref{sec:scenario1}, focuses on nonlinear system identification for two turbines operating under low turbulence conditions with varying set points. The second scenario, presented in Section~\ref{sec:scenario2}, simulates a wind farm environment with high turbulence intensity and fixed operating conditions. Finally, the robustness of the identified models is assessed through a control task, as described in Section~\ref{sec:scenario3}.

\begin{figure}[h!]
	\centering
	\begin{subfigure}[c]{0.48\textwidth} 
		\centering
		\includegraphics[width=\linewidth]{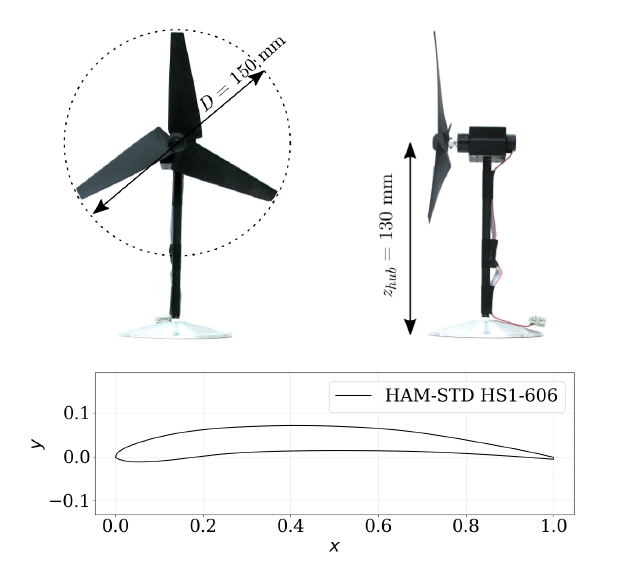}
		\caption{Wind turbine model used in the experimental campaign \citep{coudou2018experimental} and corresponding low-Reynolds-number airfoil used in the blade.}
		\label{fig:turbine}
	\end{subfigure}%
	\hfill
	\begin{subfigure}[c]{0.48\textwidth} 
		\centering
		\vspace{7mm}
		\includegraphics[width=\linewidth]{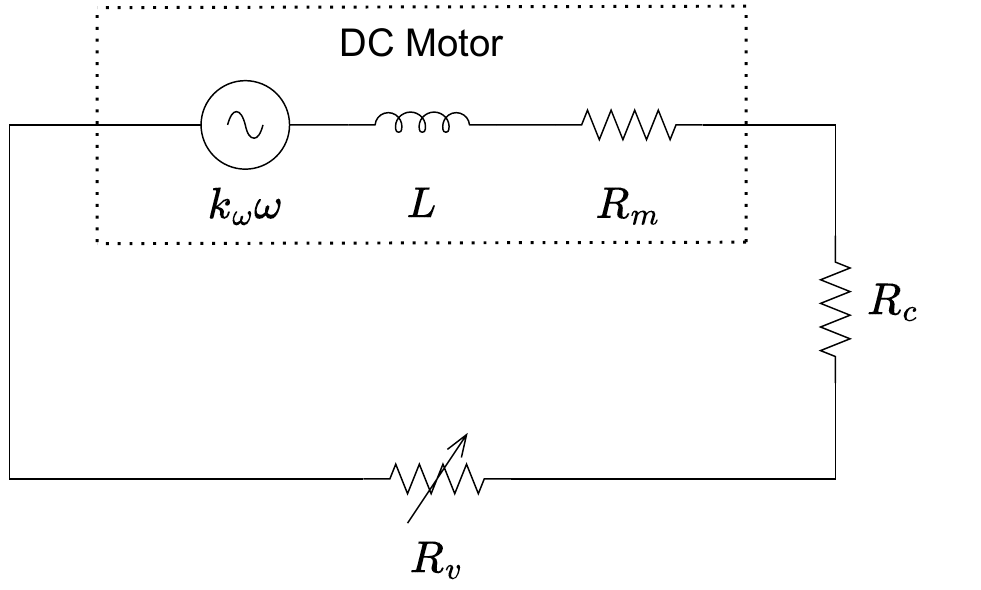}
		 \vspace{7mm}
		\caption{Electrical model of the wind turbine generator and relevant variables.}
		\label{fig:elecirc}
	\end{subfigure}
\end{figure}

\subsection{Experimental set up}
\label{sec:expsetup}

The proposed methodology was experimentally validated using three-bladed wind turbine scale models, originally designed by \citet{coudou2018experimental}. Each model features a rotor diameter of $D = 0.15\,\mathrm{m}$ and a hub height of $z_{\mathrm{hub}} = 0.13\,\mathrm{m}$. The rotor was designed based on Burton's optimal rotor theory \citep{sorensen2016optimum} and employs a low-Reynolds-number airfoil section (HAM-STD HS1-606). This design enables the turbines to achieve power and thrust coefficient performances comparable to those of full-scale turbines such as the Vestas V66-2MW offshore turbine, of which it represents a 1:440 scale version. A photograph of the turbines is shown in Figure~\ref{fig:turbine} together with the relevant dimensions and airfoil section. The total moment of inertia of the rotor has been estimated from the CAD model to be $J=2.5 \times 10^{-6}$ kg\,m$^2$.

Each rotor is directly coupled to a DC motor (Faulhaber 1331T006SR) operating as a generator. The blades are fixed hence the only actuation is through motor torque, adjusted via a variable resistance \( R_{\text{v}} \) connected to the generator. The generator torque \( \tau_g \) is modeled electrically by relating it to the current \( i \) as:
\[
\tau_g = k_{\tau} \cdot i
\]
where \( k_{\tau} \) is a motor constant. The back electromotive force (EMF) is assumed proportional to the rotor speed:
\[
V_{\text{emf}} = k_{\omega} \cdot \omega\,,
\]
with \( k_{\omega} \) also a motor-specific constant. Neglecting generator inductance \(L\) and applying mesh analysis to the equivalent circuit (Figure~\ref{fig:elecirc}), the generator torque becomes:
\begin{equation}
\tau_g = k_{\tau} \cdot \frac{k_{\omega} \omega}{R_{\text{tot}} + R_{\text{v}}}
\label{eq:elemodel}
\end{equation}
where $R_{\text{tot}} = R_{\text{m}} + R_{\text{c}}$, with $R_{\text{m}}$ and $R_{\text{c}}$ representing the internal motor and cable resistances, respectively. This relation is used in \eqref{eq:omega_dynamics} to compute the mechanical torque on the turbine shaft.

\begin{figure*}[h!]
    \centering
    \includegraphics[width=0.90\textwidth]{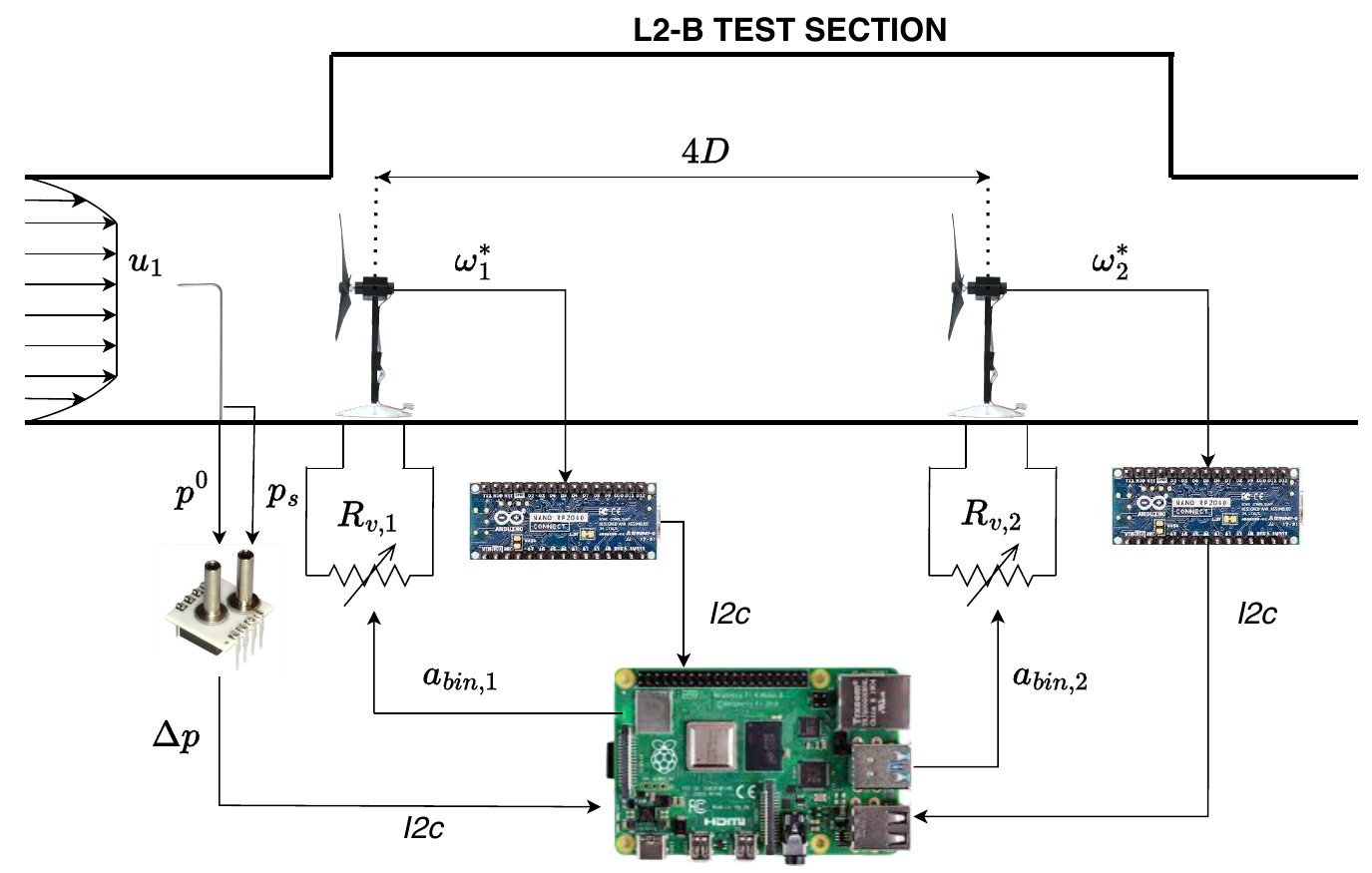}
    \caption{\label{fig:expsetup} Experimental setup in the low-turbulence scenario configuration, detailing the measurement of free-stream wind speed $u_1$, rotational speed $\omega_i^*$ of each turbine, and torque actuation via variable resistance $R_{v,i}$ controlled by a 12-bit binary signal $a_{bin,i}$.}
\end{figure*}

The experimental setup with sensors and controllers is illustrated in Figure~\ref{fig:expsetup} for the case of two turbines. A Raspberry Pi 4 is employed for real-time data acquisition and turbines' control. Each turbine is equipped with an incremental encoder for measuring the rotational speed \(\omega_i^*(t)\). These encoders feature 50 pulses per revolution (PPR), and the rotational speed is computed by an Arduino Nano through the measurement of the time interval between two consecutive pulses.

The free-stream wind speed at hub height \(u_1(z_{\text{hub}} , t)\) is measured using a Prandtl tube positioned 2 diameter upstream of the first rotor. The dynamic pressure $\Delta p = \frac{1}{2}\rho u_1^2$, defined as the difference between total $p^0$ and static pressure $p_s$, is acquired using an AMS 5812-0000-DB sensor, from which the wind speed is then derived $u_1 = \sqrt{{2\Delta  p}/{\rho}}$. A thermocouple is installed to monitor air temperature and enhance the accuracy of air density estimations.

Torque actuation is implemented by modulating the electrical load through a bank of 12 resistors, each controlled via MOSFETs. This configuration enables \(2^{12}\) discrete resistance values, ranging from 0.25~\(\Omega\) to 1024~\(\Omega\). A 12-bit binary signal \(a_{bin}\) is sent from the Raspberry Pi 4 to a shift register (SNx4HC595), which selects the active resistors. The corresponding resistance \(R_{\text{v}}\) is retrieved from a precomputed lookup table.

\subsection{Low Turbulence Scenario Identification}
\label{sec:scenario1}

The first scenario evaluates the proposed methodology for identifying the aerodynamical characteristics of two wind turbine models in a tandem configuration, where the operational conditions of both turbines are varied (Figure \ref{fig:expsetup}). This identification under varying conditions is used for the control implementation, as discussed in \ref{sec:scenario3}. The two wind turbine models are separated by four rotor diameters ($4D \approx 60$~cm). The experimental campaign was conducted in the VKI Wind Engineering facility L2-B, an open-circuit wind tunnel with a test section of dimensions $H = 0.5~\mathrm{m}$, $W = 0.5~\mathrm{m}$, and $L = 0.8~\mathrm{m}$, capable of generating uniform flow velocities up to $U_{\infty} = 35~\mathrm{m\,s^{-1}}$. The limited length of the test section prevents the development of a fully developed boundary layer representative of atmospheric conditions, resulting in low turbulence intensity. The blockage ratio is approximately 5\% \citep{coudou2021numerical}, which is low enough to avoid significant wake distortion \citep{mctavish2014experimental}.

The system identification process was performed by analyzing the response of \(\omega_i^*(t)\) to a combination of control steps in \(R_{\text{v}}\) and gradually varying freestream velocities \(u_1(t)\). The upstream turbine model was trained using seven episodes, while the downstream turbine model relied on five episodes. Each episode, as described in Section~\ref{sec:2}, corresponds to a time interval of \(T_0 = 32~\mathrm{s}\), with a data sampling frequency of \(f_s = 20~\mathrm{Hz}\). Both were then tested on four and three unseen wind/loading episodes respectively. The model identification for the two turbines was performed in separate episodes: first, the upstream turbine was trained, followed by the downstream turbine, which also utilized data from the first. This approach treats the identification as two distinct processes, even though, in principle, both turbines could be trained jointly. Figure~\ref{fig:Training_testing} illustrates the input signals used in three training episodes and two testing episodes. The left panel shows the free-stream velocity $u_1(t)$ applied during the training or testing of the upstream turbine ($u_{1,1}(t)$) and during the training or testing of the downstream turbine ($u_{1,2}(t)$). The right panel depicts the corresponding steps in generator resistance for the upstream turbine in the two cases ($R_{1,1}(t)$ and $R_{1,2}(t)$), as well as the generator resistance applied to the downstream turbine during the identification of the waked turbine model ($R_{2}(t)$).

\begin{figure*}[t!]
    \centering
    \includegraphics[width=\textwidth]{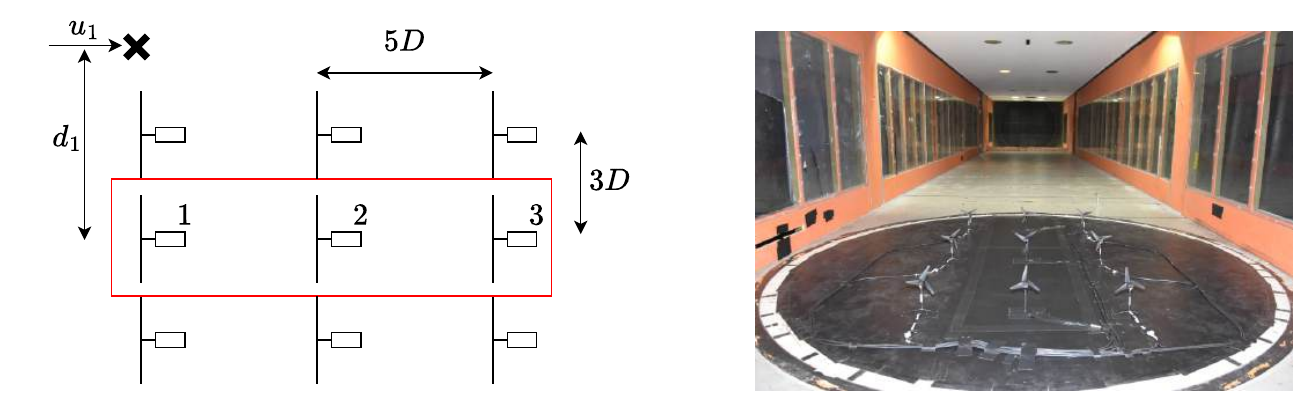}
    \caption{\label{fig:scenario1} Wind farm configuration in the VKI Wind Engineering Facility L-1B, with the three identified wind turbines highlighted in red.}
\end{figure*}

\subsection{High Turbulence Scenario Identification}
\label{sec:scenario2}

The second scenario evaluates the proposed methodology in a wind farm configuration consisting of a \(3 \times 3\) turbine array operating under high turbulence conditions. The experiments were conducted in the VKI Wind Engineering facility L1-B. This is a low-speed, closed-loop wind tunnel with a test section measuring $H = 2~\text{m}$, $W = 3~\text{m}$, and $L = 20~\text{m}$, delivering a maximum wind speed of $U_{\infty} = 60~\text{m/s}$. The length of the test section allows the development of a fully established turbulent boundary layer at the location of the wind farm, which is positioned at the downstream end of the tunnel and thus entirely within the boundary layer. Further details on how the boundary layer generated in the L1-B tunnel replicates atmospheric flow conditions can be found in \citep{conan2012wind}.

A schematic of the experimental setup is shown in Figure~\ref{fig:scenario1}. The turbine models were arranged with spanwise and streamwise spacings of 3D and 5D, respectively.  For the nonlinear identification test, three turbines from the central row of the array were selected: one located in the freestream and two positioned in the wake region. This setup targets turbines operating near their optimal performance point under steady turbulent inflow, consistent with typical wind turbine control strategies. Notably, the freestream wind velocity was not measured directly at the first turbine. Instead, as shown in Figure~\ref{fig:scenario1}, a Prandtl tube was placed $d_1 = 1.2\,\mathrm{m}$ upstream in the streamwise direction, aligned with the first row of turbines in the spanwise direction.

To maintain a consistent operating point, the resistance \(R_{\text{var}}\) connected to the DC generator was fixed at \(R_{\text{val}} = 1~\Omega\), corresponding to a tip-speed ratio of \(\lambda_1 = 4.5\) for a reference inflow velocity of \(u_1 \approx 8.5~\text{m/s}\). This value approximately maximizes the power coefficient, as shown in \cite{coudou2021numerical}. The experimental setup and measurement procedure matched those described in previous sections. All signals were sampled at \(f_s = 20~\text{Hz}\) over episodes lasting approximately \(T_0 = 20~\text{s}\), with 80\% of each dataset used for training and the remaining 20\% reserved for testing.

\begin{figure*}[htbp]
	\centering
	\renewcommand{\thesubfigure}{\alph{subfigure}} 
	
	\begin{subfigure}[b]{0.48\textwidth}
		\centering
		\includegraphics[width=\linewidth]{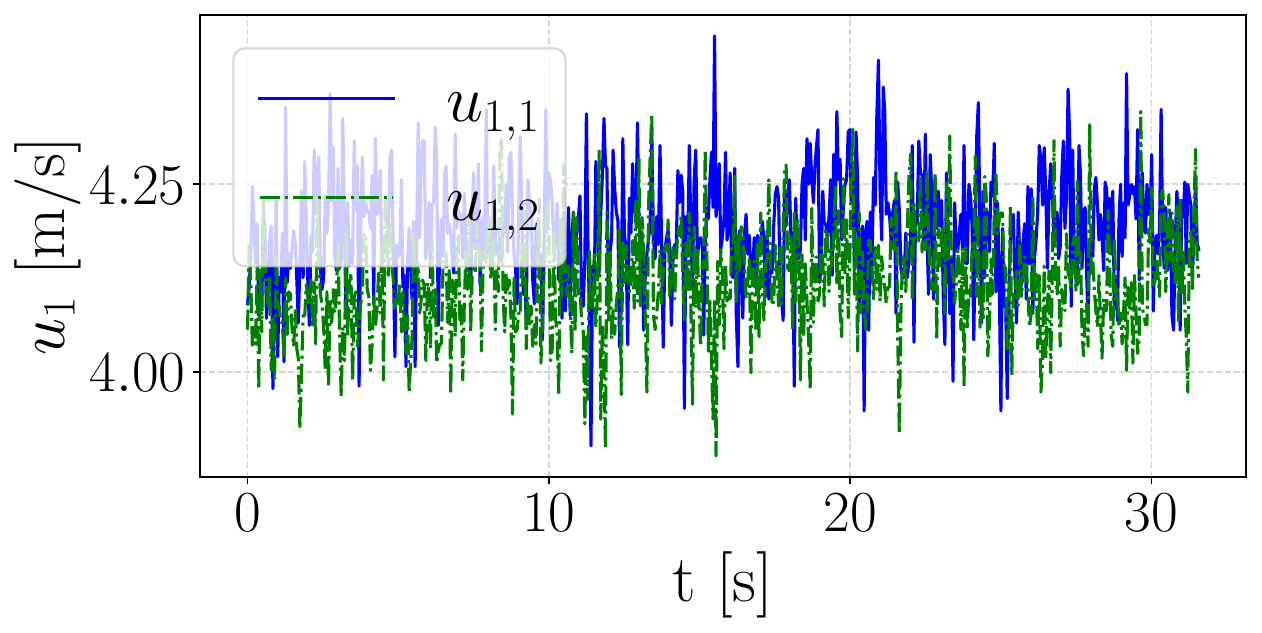}
	\end{subfigure}%
	\hfill
	\begin{subfigure}[b]{0.48\textwidth}
		\centering
		\includegraphics[width=\linewidth]{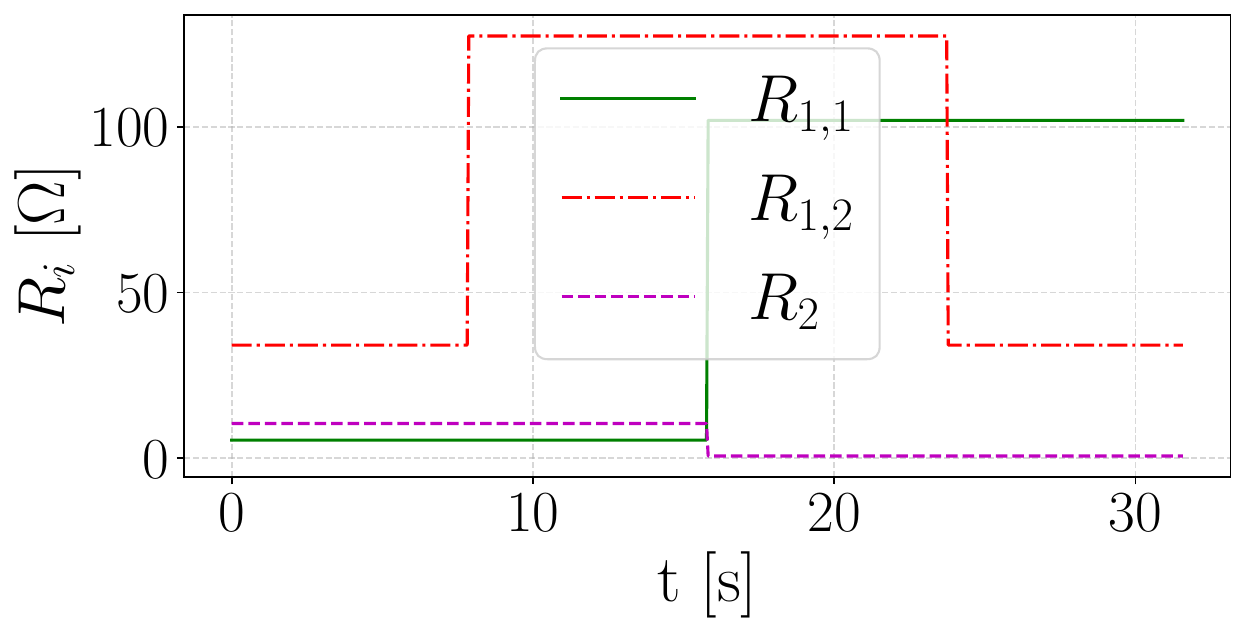}
	\end{subfigure}
	
	\vspace{1mm}
	
	\begin{subfigure}[b]{0.48\textwidth}
		\centering
		\includegraphics[width=\linewidth]{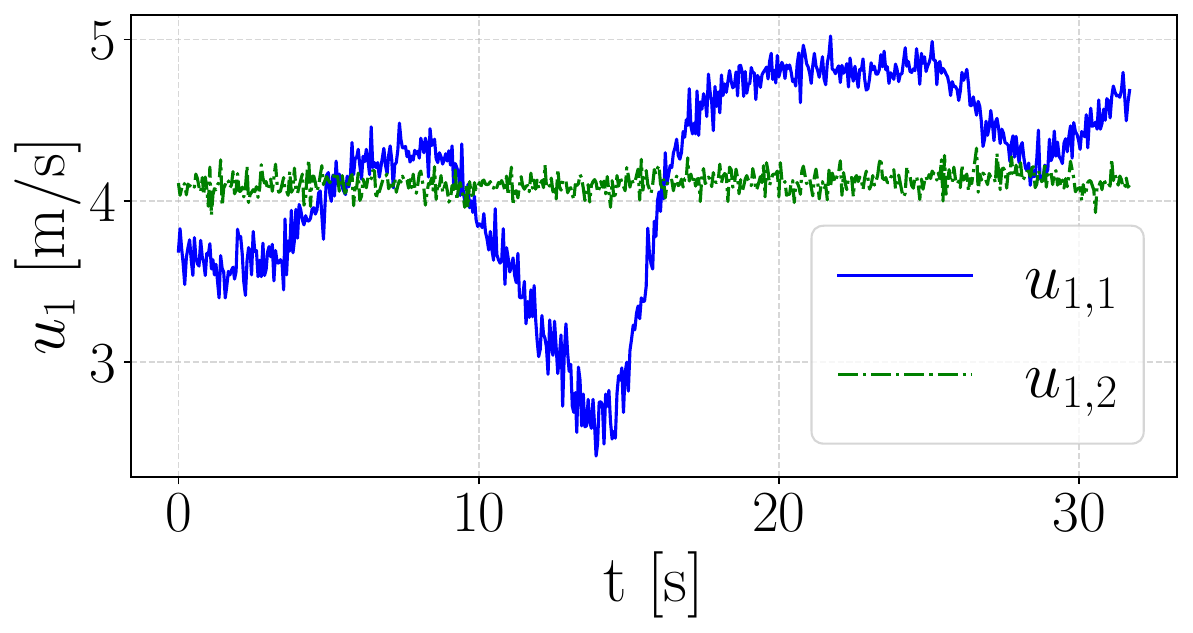}
	\end{subfigure}%
	\hfill
	\begin{subfigure}[b]{0.48\textwidth}
		\centering
		\includegraphics[width=\linewidth]{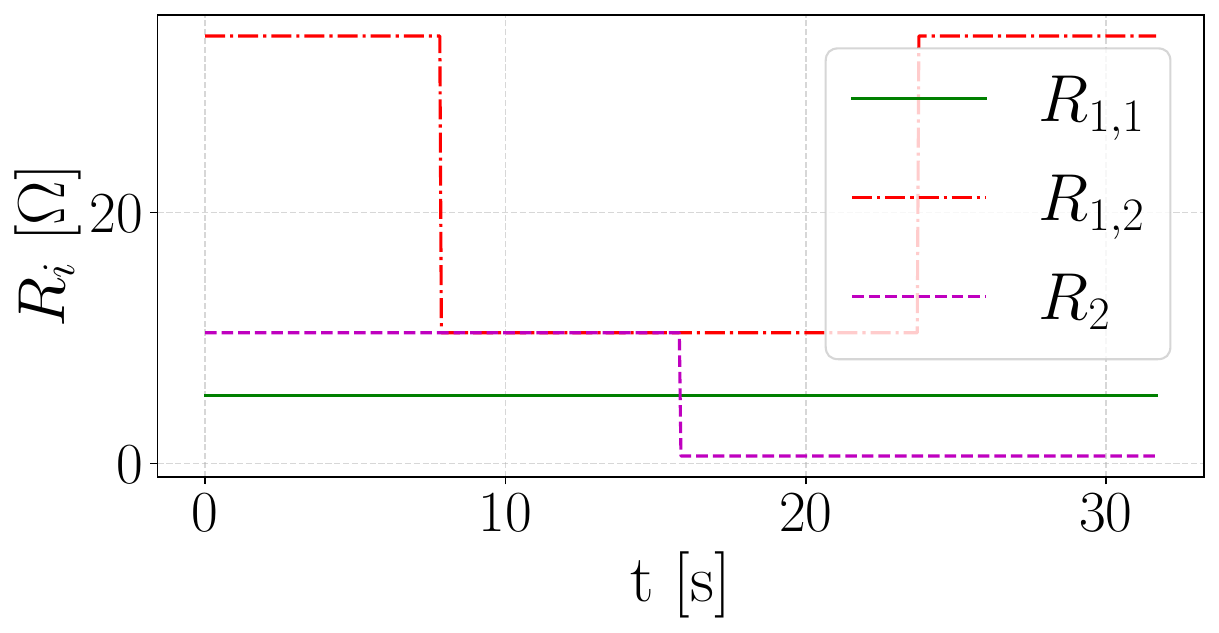}
	\end{subfigure}
	
	\vspace{1mm}
	
	\begin{subfigure}[b]{0.48\textwidth}
		\centering
		\includegraphics[width=\linewidth]{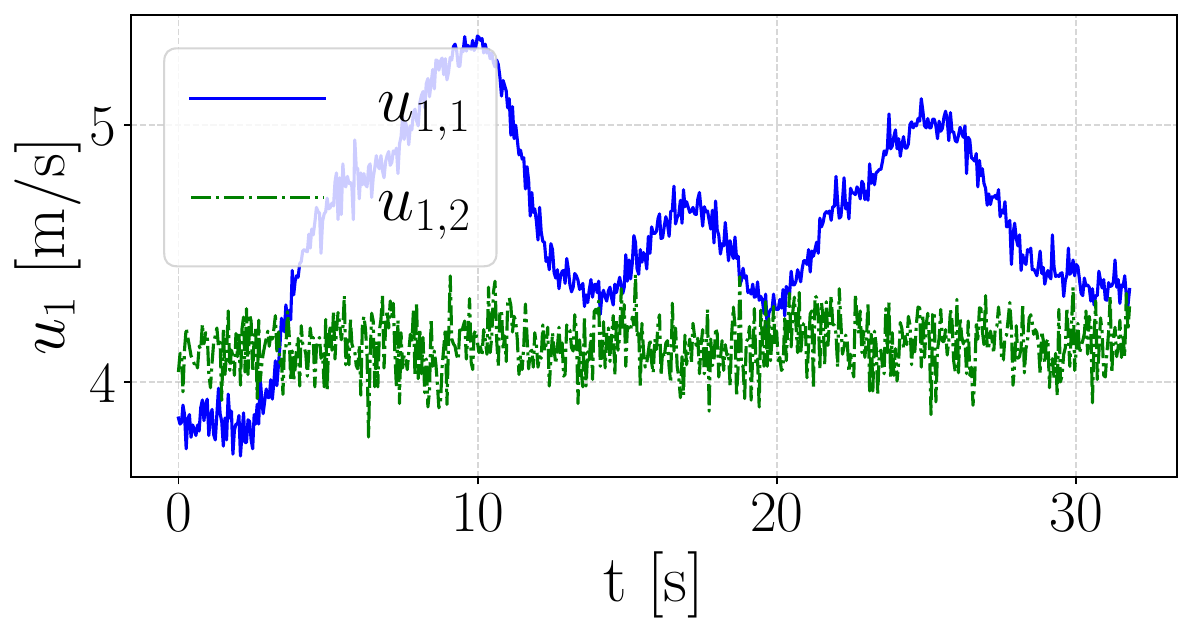}
	\end{subfigure}%
	\hfill
	\begin{subfigure}[b]{0.48\textwidth}
		\centering
		\includegraphics[width=\linewidth]{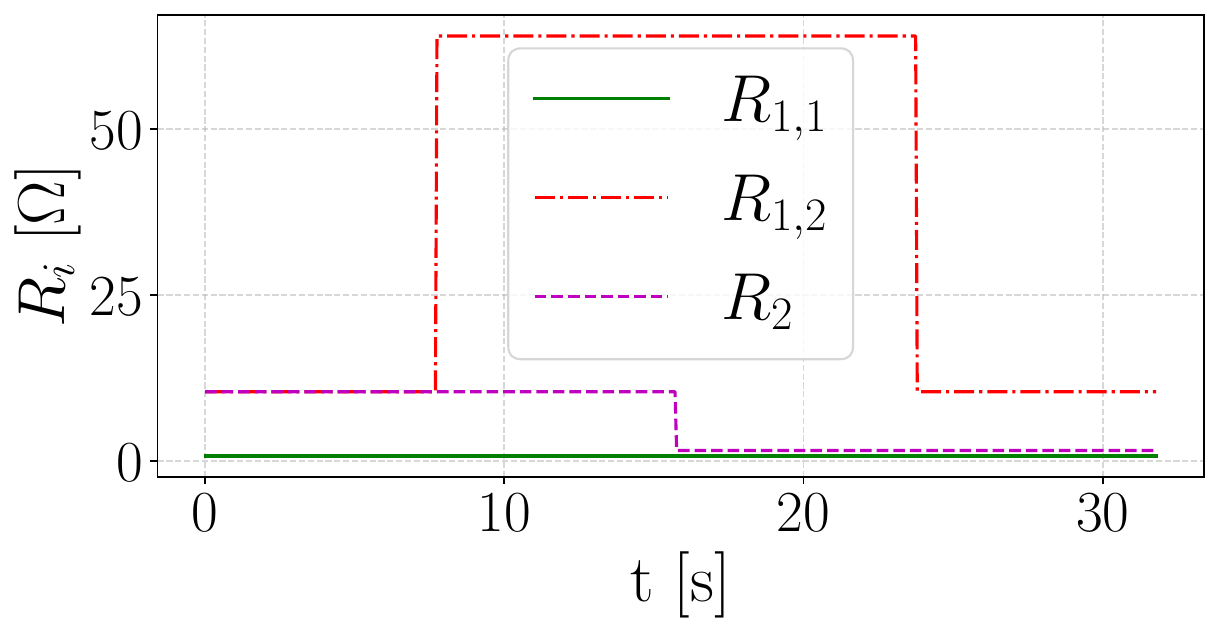}
	\end{subfigure}
	
	\vspace{1mm}
	
	\begin{subfigure}[b]{0.48\textwidth}
		\centering
		\includegraphics[width=\linewidth]{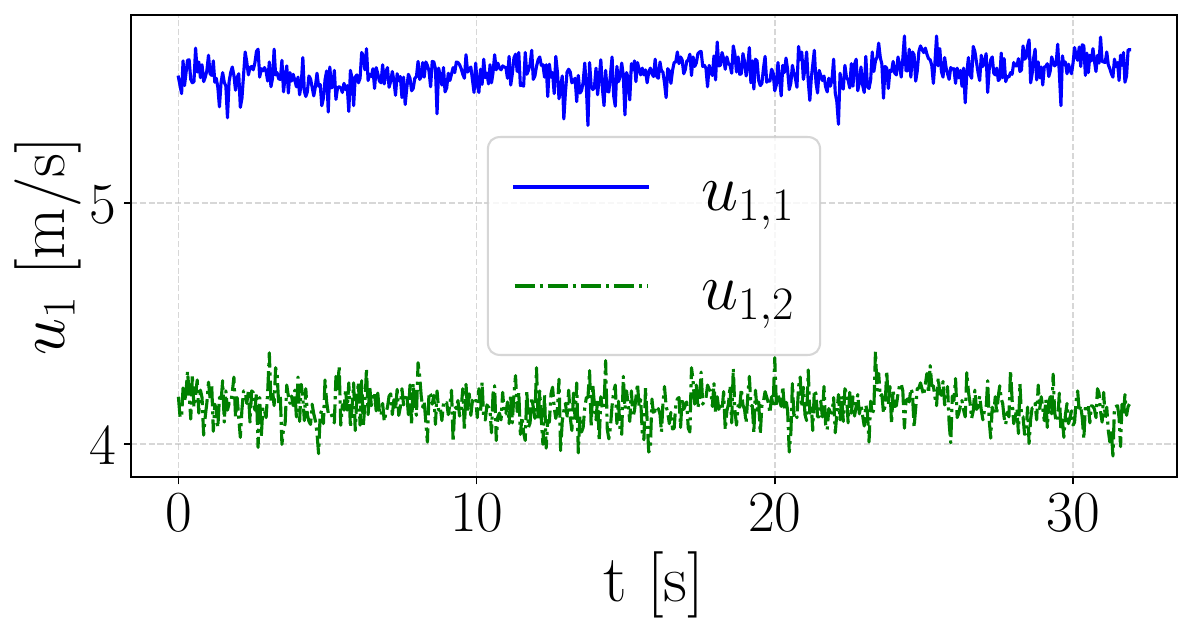}
	\end{subfigure}%
	\hfill
	\begin{subfigure}[b]{0.48\textwidth}
		\centering
		\includegraphics[width=\linewidth]{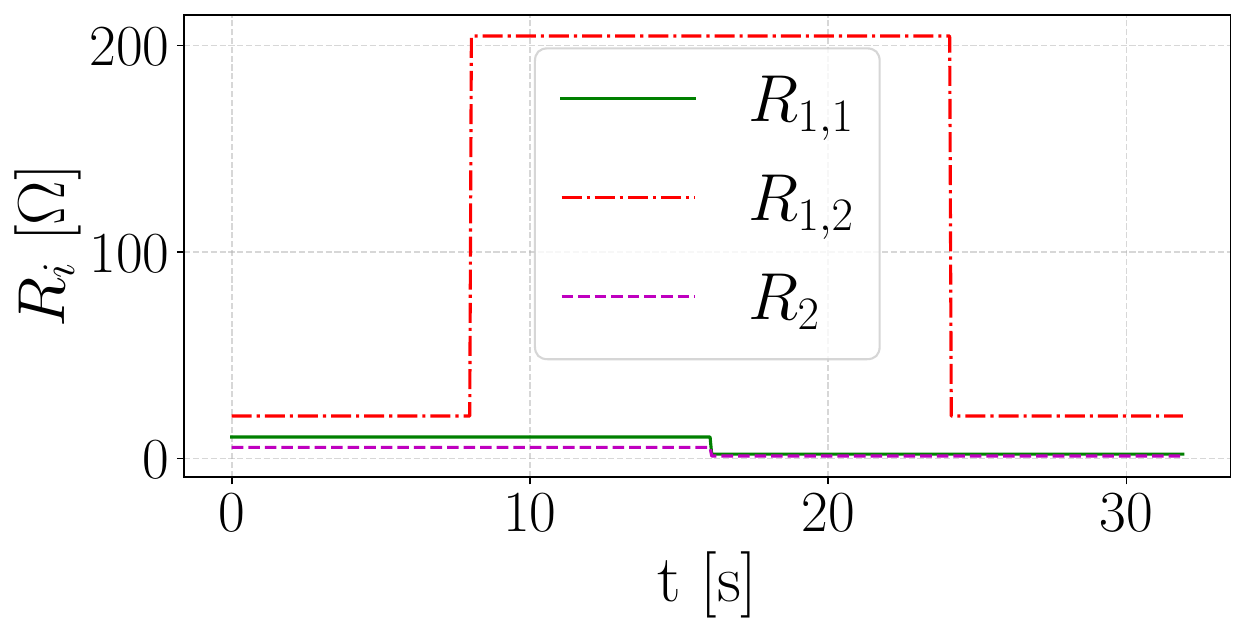}
	\end{subfigure}
	
	\vspace{1mm}
	
	\begin{subfigure}[b]{0.48\textwidth}
		\centering
		\includegraphics[width=\linewidth]{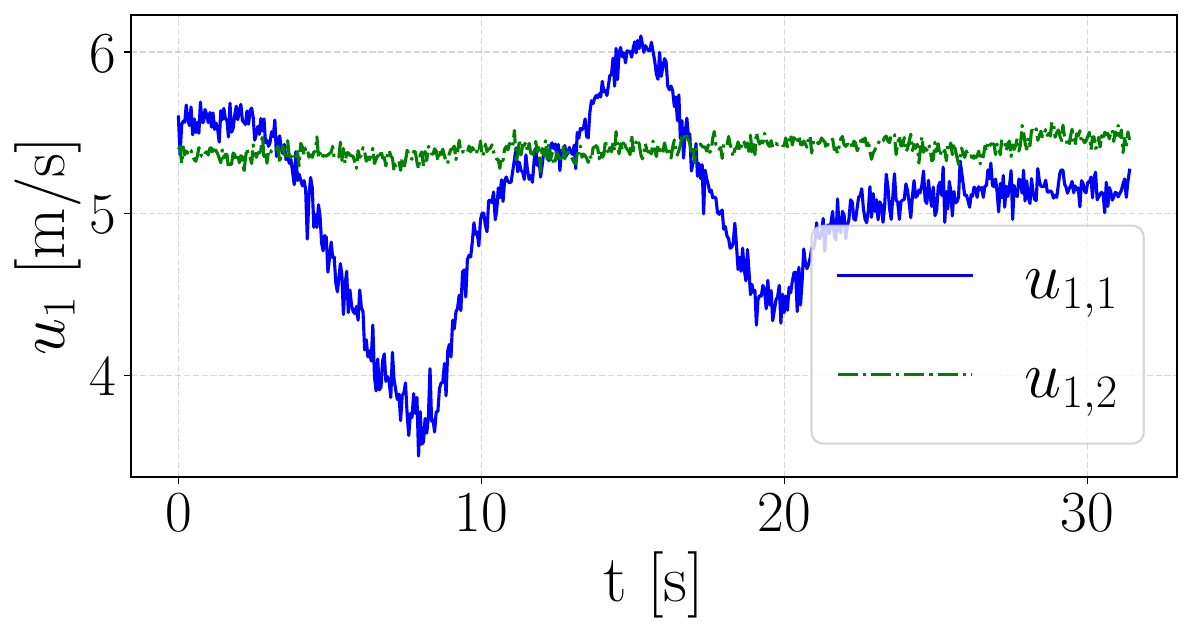}
	\end{subfigure}%
	\hfill
	\begin{subfigure}[b]{0.48\textwidth}
		\centering
		\includegraphics[width=\linewidth]{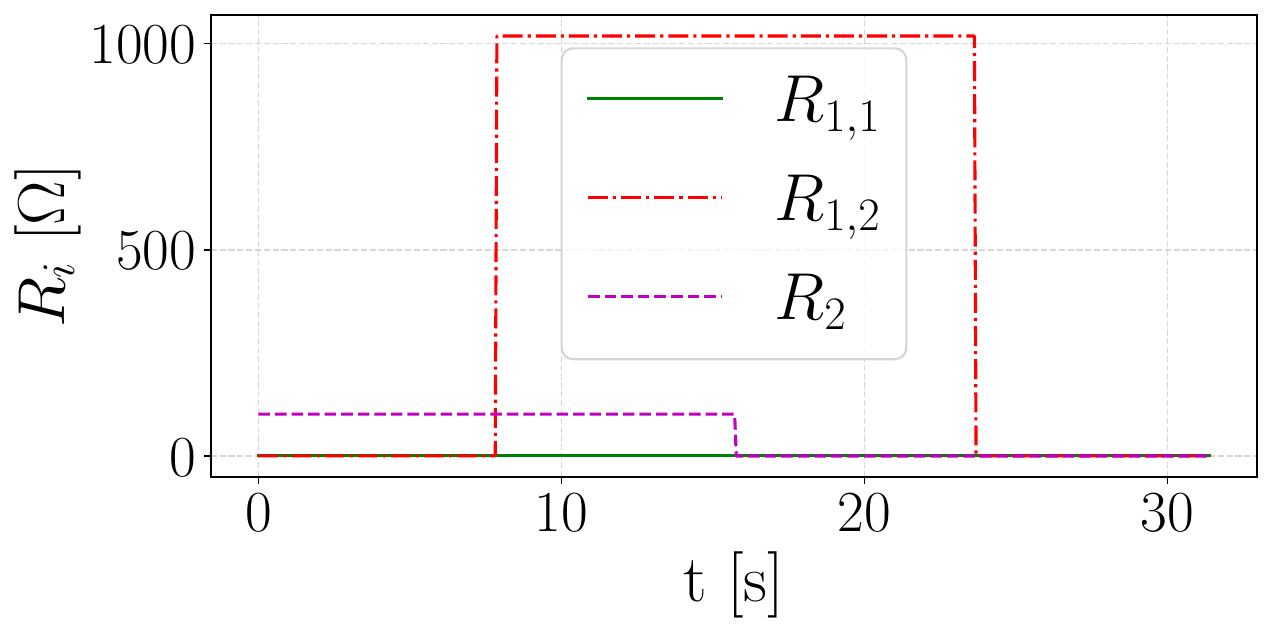}
	\end{subfigure}
	
	\caption{Free stream velocity (left) and generator resistance evolution (right) for three representative training episodes (first three rows) and two testing episodes for the low turbulence scenario described in Section \ref{sec:scenario1}.}
	\label{fig:Training_testing}
\end{figure*}

\subsection{Control under Low Turbulence Conditions}
\label{sec:scenario3}

To evaluate the accuracy of the identified models for control purposes, we return to the tandem configuration in section \ref{sec:scenario2}. For both turbines, we consider the classic \(K\omega^2\) law, commonly used in Region 2 to maximize power extraction \citep{pao2009tutorial}. This controller gives a quadratic torque law

\begin{equation}
\label{Komegasquare}
\tau_g = K\omega^2\,,
\end{equation}
where the constant \(K\) is defined as (\citet{pao2009tutorial}):
\begin{equation}
\label{K_eq}
K = \frac{1}{2} \rho A R^3 \frac{C_{P_{\text{max}}}}{\lambda_{\text{max}}^3}\,,
\end{equation} where \(\lambda_{\text{max}}\) denotes the tip-speed ratio corresponding to the maximum power coefficient \(C_{P_{\text{max}}}\).

In this work, the \(K\omega^2\) controller is employed in terms of electrical generator resistance. Combining \eqref{Komegasquare} and \eqref{K_eq} with \eqref{eq:elemodel} gives 

\begin{equation}
\label{eq:control}
R_{\text{v}} = \frac{k_\tau k_\omega}{K \omega} - R_{tot}\,.
\end{equation}

This model-based control strategy offers a valid test to assess the fidelity of the assimilated dynamic model, as the controller performance is inherently tied to the accuracy of the assumed \( C_P(\lambda, Re) \) distribution. Moreover, in the investigated scenario, the controller is not used under typical operational assumptions that is to track the tip-speed ratio maximizing power production. Instead, the setpoint in terms of tip-speed ratio is intentionally varied to stress-test the model across a wider operating range.

To ensure reliability of the sensor measurements, all input signals are filtered using a first-order Butterworth low-pass filter with a cutoff frequency of \(f_{\text{c.o.}} = 2\)~Hz. The performance of the controller is evaluated using three different modeling approaches: (1) the model identified from data, (2) the parametric model obtained from steady-state data regression \eqref{C1} which offered the initial condition for the adjoint-based refinement, and (3) a model based on regression (following a procedure similar to that described in Section \ref{sec:3.1}) of Blade Element Momentum (BEM) simulation results. The motivation for using a BEM-based controller is that it is the standard tool for obtaining aerodynamic models for control in wind turbines. The BEM simulations were set up by first computing the aerodynamic performance of the airfoil in terms of lift and drag coefficients as a function of the angle of attack, using XFOIL at different Reynolds numbers. These aerodynamic polars were then used to compute the power coefficient through BEM simulations, implemented using the `CcBlade` module from the WISDEM Python package \citep{ning2013ccblade}. This comprehensive evaluation enables a comparative analysis of controller performance under different modeling assumptions.

\section{Results}\label{sec:5}

The results are grouped into three subsections. First, Section \ref{sec6p1} presents the system identification results for the wake tandem configuration in low turbulence conditions described in Sec. \ref{sec:scenario1}. Section \ref{sec6p2} then moves to the identification results in the high turbulence scenario described in Sec. \ref{sec:scenario2}. Finally Section \ref{sec6p3} test the accuracy of the identified models for model based control purposes.

\begin{figure*}[h!]
	\centering
	\includegraphics[width=\textwidth]{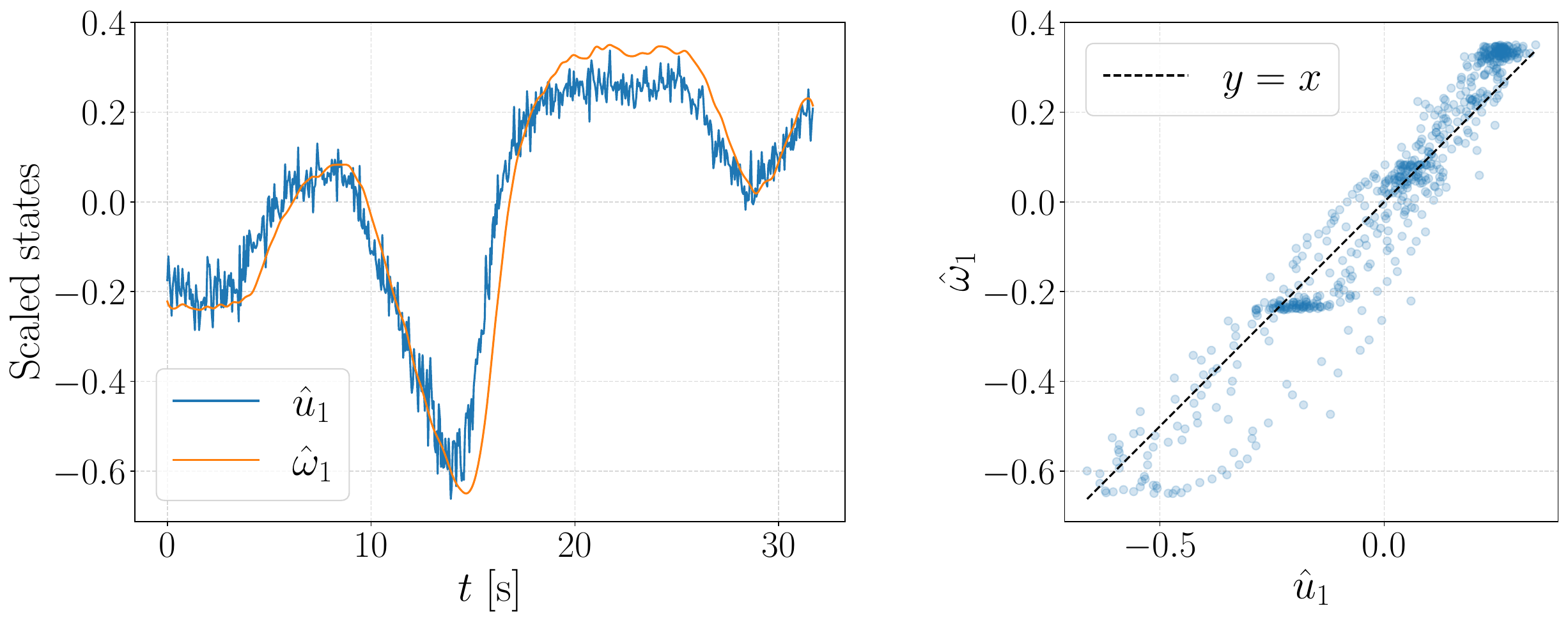}
	\caption{\label{fig:Correlation_1}
		\textbf{Left:} Time evolution of the hub-height wind velocity of the first turbine (\(\hat{u}_1\)) and its angular velocity (\(\hat{\omega}_1\)) for an example test case. Both signals are normalized by mean-centering and min-max scaling. \textbf{Right:} Scatter plot showing the correlation between the two signals. Data are from the ``low-turbulence'' scenario described in Section~\ref{sec:scenario1}.}
\end{figure*}

\begin{figure*}[h!]
	\centering
	\includegraphics[width=\textwidth]{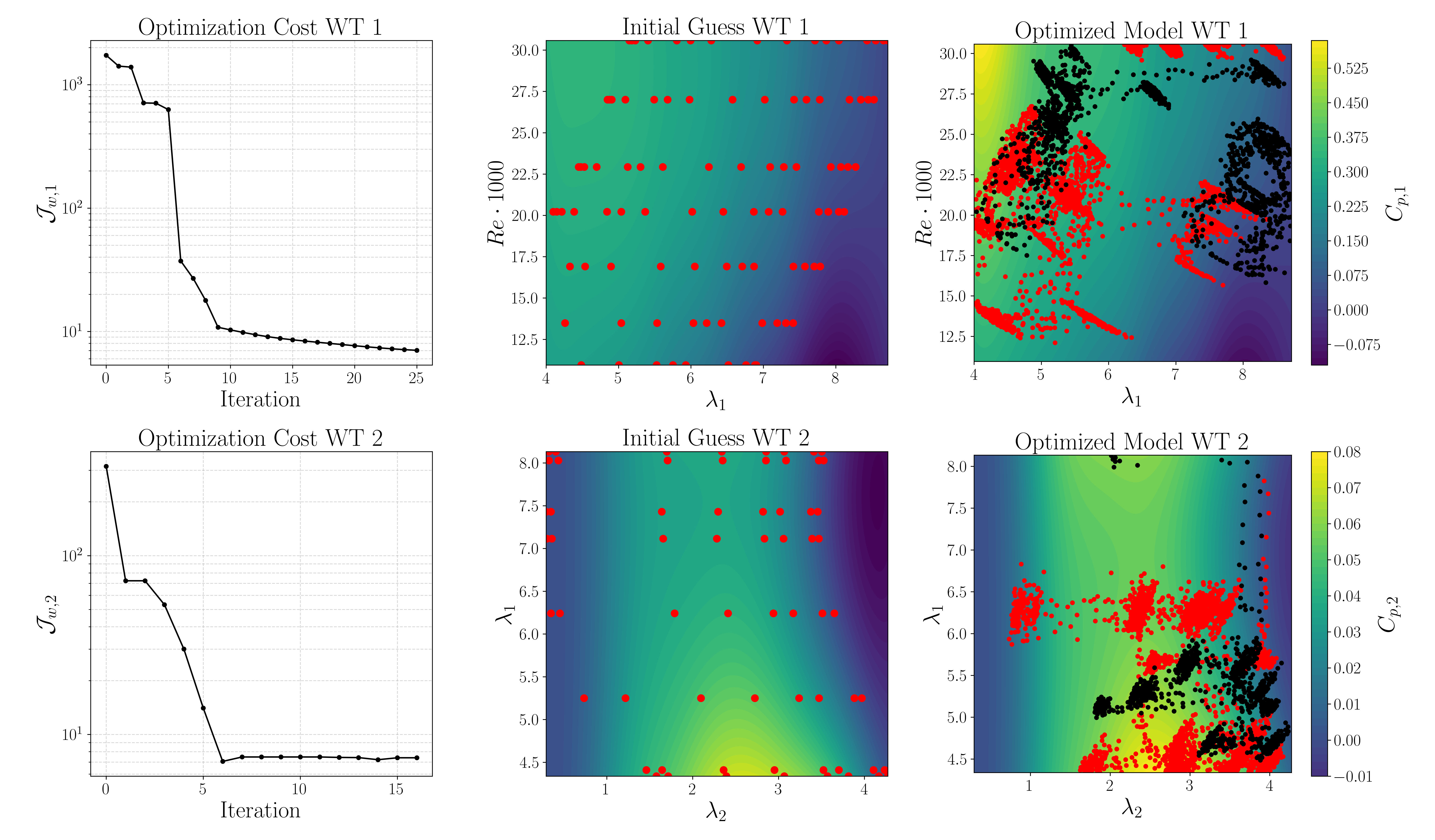}
	\caption{ Evolution of the optimization cost functions (left), initial guesses and training data (center), and optimized models (right) for the power coefficients of the upstream (top row) and downstream (bottom row) wind turbines. The red markers in the center plots represent the steady-state training data points used to obtain the initial guesses for the optimization, whereas in the right plots they correspond to trajectories encountered during the training test cases. Similarly, the black markers represent trajectories encountered during the testing cases.}
	\label{fig:CPcomparison}
\end{figure*}

\subsection{Identification in Low Turbulence}\label{sec6p1}

In this scenario, turbulence fluctuations are small compared to the large-scale variations in the incoming wind velocity. Additionally, the wind velocity is measured at hub height, consistent with the definitions in Figure~\ref{fig1}. As a result, the nonlinear system identification is performed in ideal conditions, where the model in \eqref{eq:omega_dynamics} provides a reliable instantaneous relationship between the $\omega_1(t)$ and $u_1(t)$ because these variables are reasonably correlated. This is only true if the torque on the generator of the first turbine remains constant during the episode; otherwise, changes in torque alter the rotational speed independently of the wind speed. Figure \ref{fig:Correlation_1} shows the time series of these signals, normalized with mean-centering and min-max normalization (scaled quantities denoted with hat) on the left, and a plot showing their strong correlation on the right. 

Figure \ref{fig:CPcomparison} illustrates, on the left, the evolution of the cost function over the optimization iterations for both the upstream (first row) and downstream (second row) wind turbines using all the training episodes. The optimization clearly reduces both cost functions, thus improving the initial guesses from \eqref{C1}. Convergence is achieved within a relatively small number of episodes, highlighting the efficiency of the ADAM optimizer with the adjoint-based gradient computation. In the center, Figure \ref{fig:CPcomparison} shows the power coefficient curves obtained using the initial guesses for the optimization derived from \eqref{C1}, based on training grids of $n_{\lambda_1}\times n_{Re}$ and $n_{\lambda_2}\times n_{\lambda_1}$ samples (indicated by red markers) in the $(\lambda_1\text{–}Re)$ and $(\lambda_2\text{–}\lambda_1)$ planes (as described in Section \ref{sec:3.1}). These training samples were selected to cover, as broadly as possible, the physical operating space of both wind turbines.

\begin{figure}[H]
	\centering
	\includegraphics[width=\textwidth]{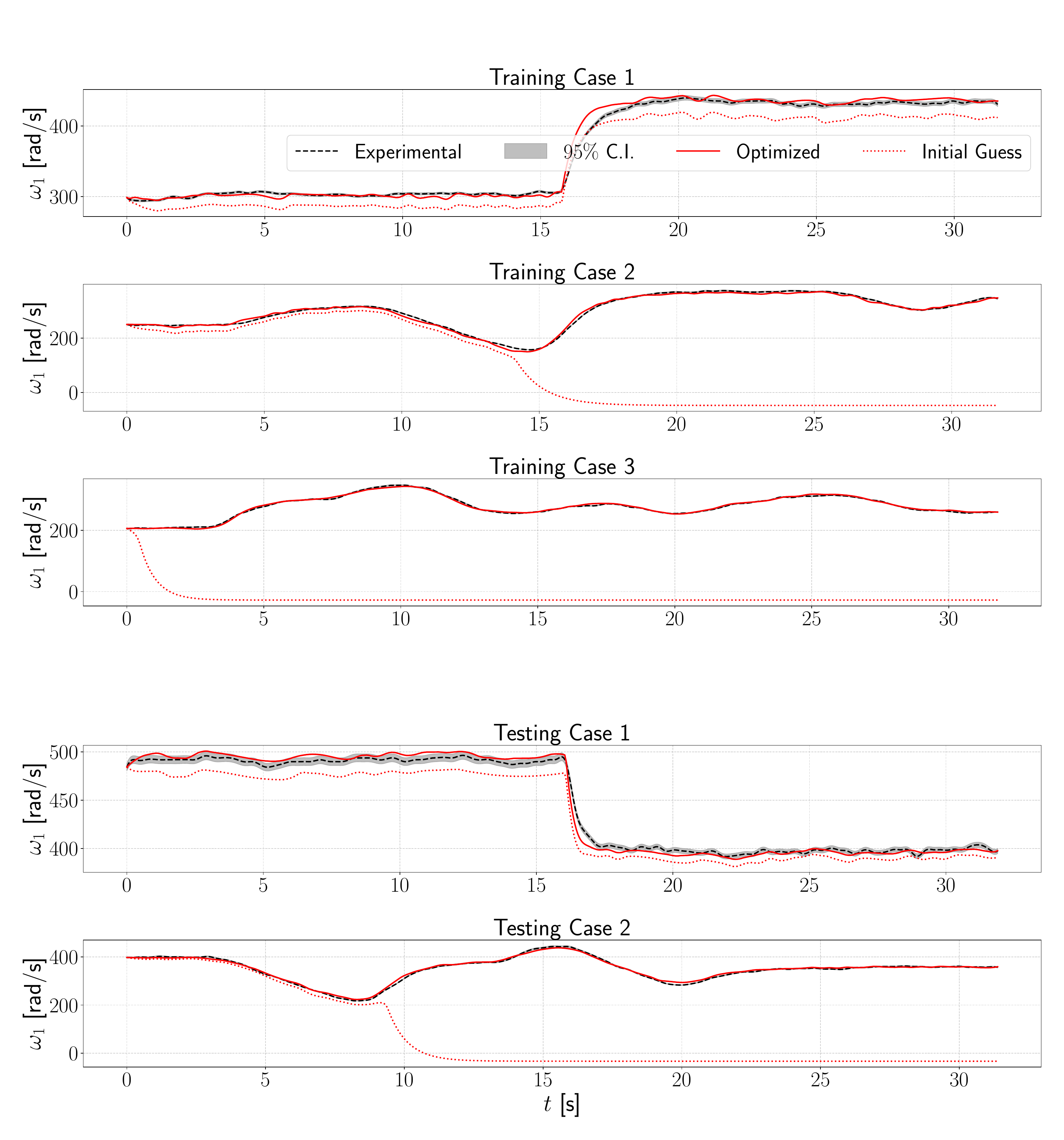}
	\caption{\label{fig:comparisontra} Time evolution of the measured rotational speed of the upstream wind turbine, $\omega_1^*(t)$ (black line with 95\% confidence interval in gray), compared with the predictions from the adaptive models. The solid red line represents the prediction using the optimal solution ($\omega_1(t; \mathbf{w}_1)$), while the red dotted line shows the prediction from the initial guess model ($\omega_1(t; \mathbf{w}_{1,0})$).}
\end{figure}

On the right, Figure~\ref{fig:CPcomparison} presents the final 2D regressions of the power coefficients: for the first turbine (top), as a function of tip-speed ratio and Reynolds number, and for the second turbine (bottom), as a function of the two tip-speed ratios. The scatter points show the states visited in the training dataset (red markers) and in the testing dataset (black markers). Overall, these provide good coverage, except in the region of low $\lambda_1$ and high $Re$ for the first turbine, and in the regions of (i) high $\lambda_1$ and (ii) simultaneously low $\lambda_1$ and low $\lambda_2$ for the second turbine. For the upstream turbine, this gap in the training data results in a nonphysical model in the unsampled region, highlighting the importance of covering a broad range of operating conditions during training. However, considering that the physical system is unable to reach these conditions (e.g., $\lambda_1\approx 4$ with $Re\approx 30000$) because of physical limitations, the lack of model accuracy in this area is inconsequential for both predictive and control purposes. It is worth stressing that the optimized curves do not necessarily reflect the true aerodynamic characteristics of the turbines but rather serve as surrogate models that best reproduce the experimental data, potentially absorbing effects such as measurement noise, generator model errors, or parameter uncertainties (e.g., from CAD estimates).

\begin{figure}[H]
	\centering
	\includegraphics[width=\textwidth]{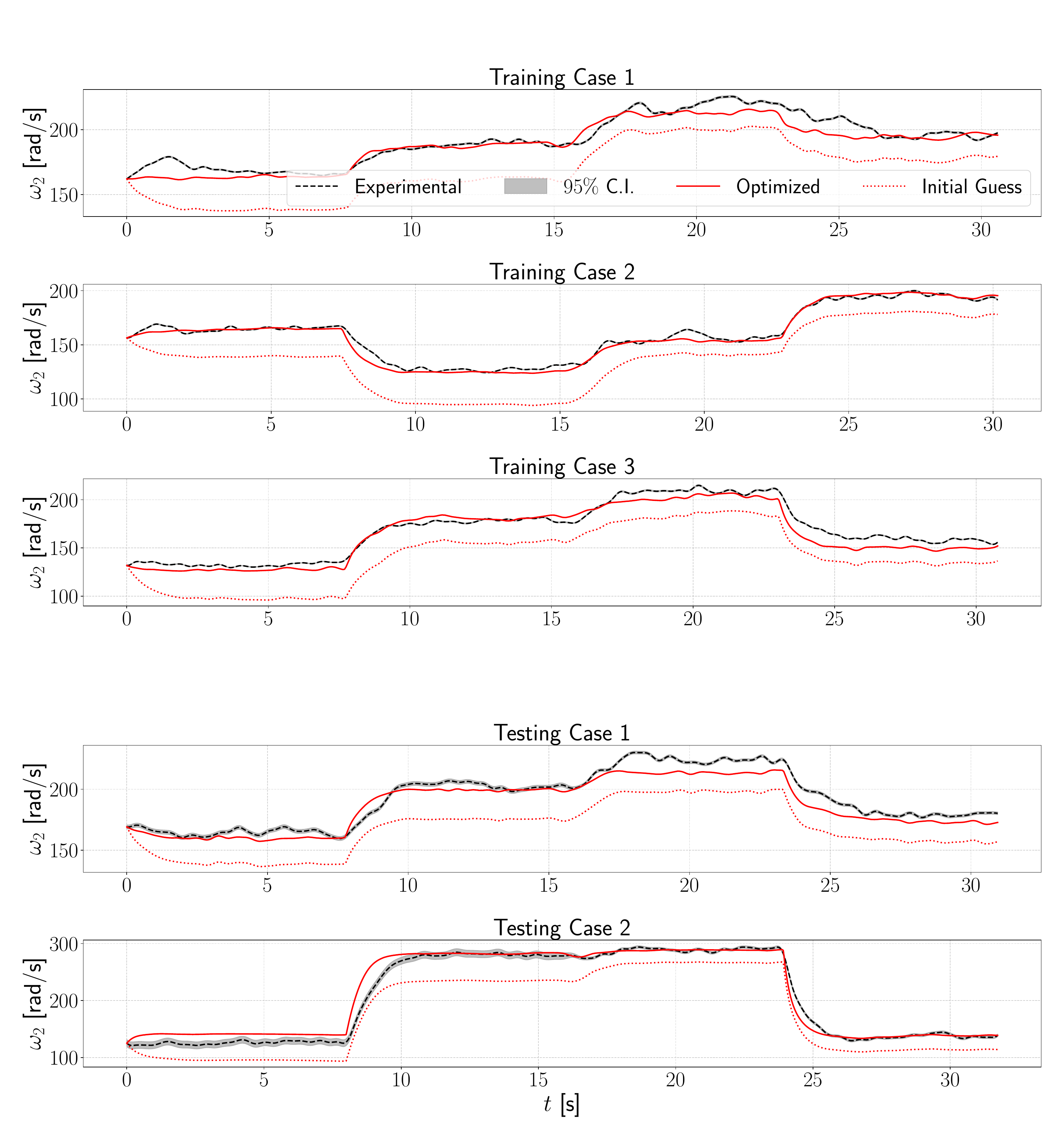}
	\caption{\label{fig:comparisontra2} 
		Time evolution of the measured rotational speed of the downstream wind turbine, $\omega_2^*(t)$ (black line with 95\% confidence interval in gray), compared with the predictions from the adaptive models. The solid red line represents the prediction using the optimal solution ($\omega_2(t; \mathbf{w}_2)$), while the red dotted line shows the prediction from the initial guess model ($\omega_2(t; \mathbf{w}_{2,0})$).}
\end{figure}

Figures \ref{fig:comparisontra} and \ref{fig:comparisontra2} show the time evolution of the measured rotational speed of the wind turbines, $\omega_i^*(t)$, represented in black, along with the associated 95\% confidence interval (gray shaded area), and the corresponding predictions of the adaptive models. For both the upstream and downstream turbines, we report the model prediction obtained using the optimal solution, with continuous line, and the model prediction using the initial guess for the model coefficients with the dashed lines. Each figure presents three randomly selected training trajectories and two test trajectory for both turbines, selected from the pool of available training and testing datasets. The adaptive model demonstrates good agreement with the measured trajectories for both turbines, although predictive performance are slightly lower for the waked turbine, particularly in the description of fluctuation at higher frequencies. This can be attributed to two main factors: (1) reduced coherence between the signals at higher frequencies, and (2) the low-pass filtering effect of the identification procedure, which prioritizes the reconstruction of larger-scale variations.

Moreover, the comparison between the initial and optimized models shows that the main improvement lies in a significantly better representation of the system's stability properties. In some cases (Training Cases 2, 3, and Testing Case 2 for the upstream turbine), the initial parameter guess from \eqref{C1} leads the model to be undesirably and unrealistically attracted to the fixed point $\omega=0$, resulting in shutdown-like behavior that does not match the collected data (see also Section~\ref{sec3p2}). This does not imply, however, that imposing the constraint in \eqref{eq:stability} lacks merit. On the contrary, incorporating stability constraints could improve model robustness and restrict the search space during optimization. These aspects will be explored in future works.

On a practical note, we emphasize that the cost function exhibits significant stiffness in regions of the parameter space where changes in stability properties occur. To help the ADAM optimizer adapt to these abrupt changes, we reinitialize it whenever a threshold is crossed (\( \Delta \mathcal{J}_{w,i} > 100 \)), effectively restarting the gradient smoothing procedure in \eqref{grad_SMOOTH}. This re-initialization occurs, for instance, between iterations 3–4 and 6-7 in the upstream turbine case, as shown in the cost function evolution in Figure~\ref{fig:CPcomparison}. Additionally, we observe that once the cost function falls below a small threshold (\( \mathcal{J}_{w,i} < 10 \)), the momentum term becomes detrimental to convergence. In such cases, reducing the learning rate by a factor of 10 improves convergence by enabling finer adjustments in promising regions of the parameter space.

\subsection{Identification in High Turbulence}\label{sec6p2}

\begin{figure*}[h!]
	\centering
	\includegraphics[width=\textwidth]{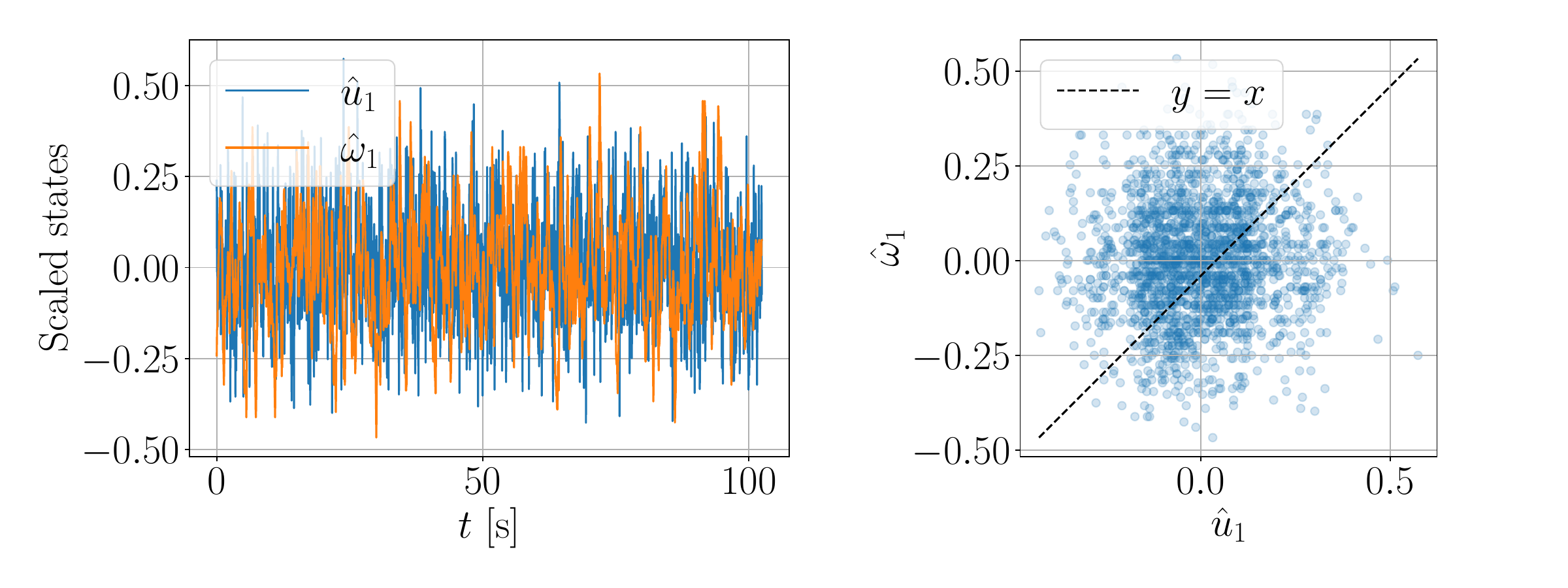}
	\caption{\textbf{Left:} Time evolution of the dimensionless wind speed at the location highlighted in Figure~\ref{fig:scenario1} ($\hat{u}_1$) and the dimensionless rotational speed of the first turbine ($\hat{\omega}_1$). \textbf{Right:} Scatter plot showing the low correlation inflow conditions and turbine dynamics.}
	\label{fig:coherency2}
\end{figure*}

\begin{figure*}[h!]
	\centering
	\includegraphics[width=\textwidth]{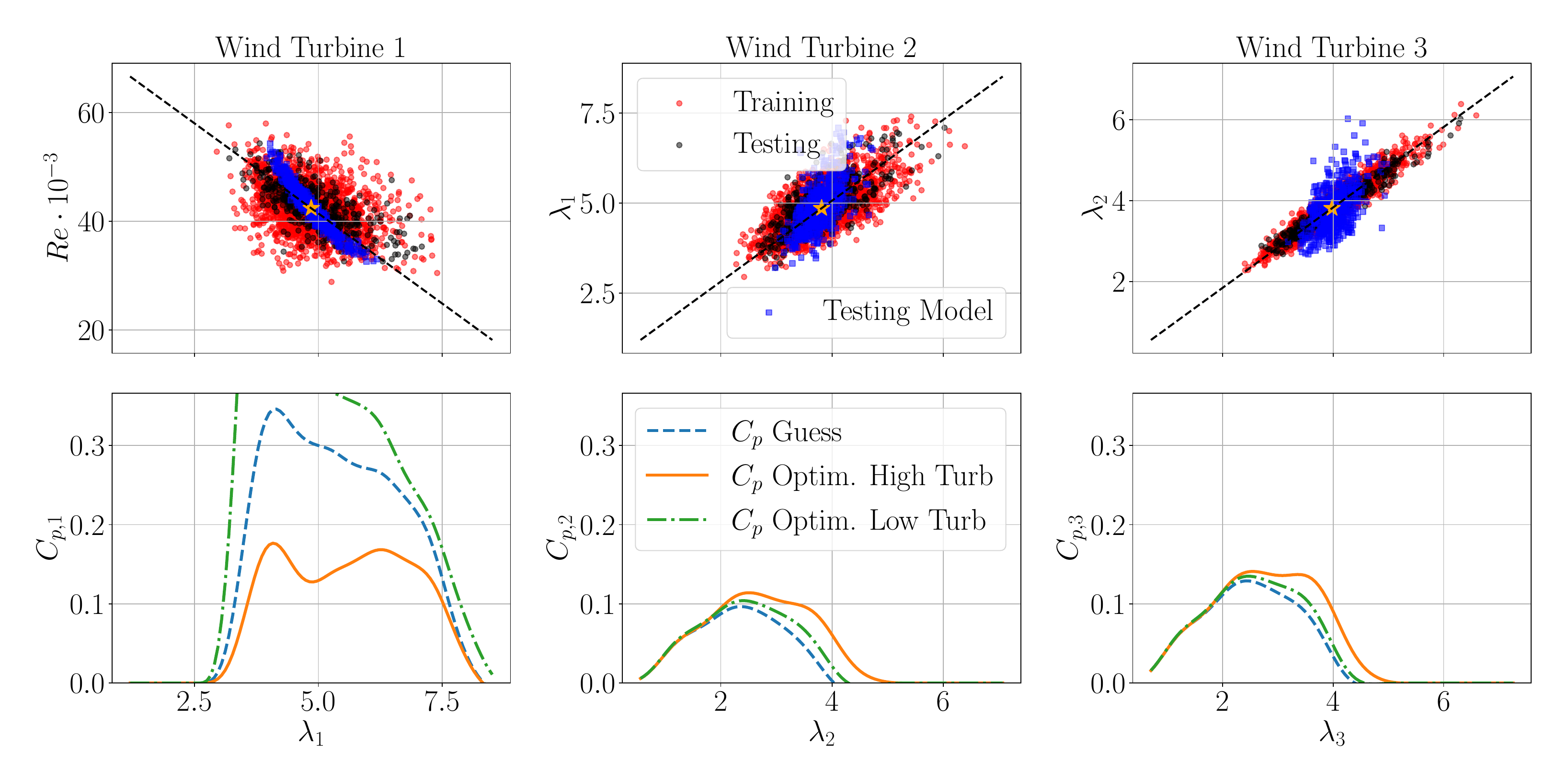}
	\caption{ Summary of the identification result for the wind farm scenario under highly turbulent conditions. The first row shows the location of the sampling and testing points for the three turbines (ref. to Figure \ref{fig:scenario1} for their position). The second row shows the power coefficient curves obtained from the initial guess and the optimized weights for the high-turbulence scenario, as well as those corresponding to the low-turbulence scenario.}
	\label{fig:CPcomparison2}
\end{figure*}

As in the previous section, we begin by analyzing the correlation between the normalized incoming velocity $\hat{u}_1(t)$ and the rotational speed of the upstream turbine $\hat{\omega}_1(t)$, shown in Figure \ref{fig:coherency2}. This test case corresponds to stationary conditions with high turbulence, where wind speed fluctuates between 7 and 11 m/s. As a consequence that the velocity measurement is not taken at the hub of the turbine, the variables $u_1(t)$ and $\omega_1(t)$ exhibit no meaningful correlation, meaning that it impossible to reconstruct the turbine’s rotational speed time series as done in the previous case. Nonetheless, the objective here is to capture the time-averaged signal, even in the absence of short-term correlation. This average remains indicative of incoming fluctuations occurring at spatial and temporal scales much larger than the duration of the episode through which the identification is carried out, and it can thus still provide valuable information for control strategies at the wind farm level.

\begin{figure}[h!]
	\centering
	\includegraphics[width=\textwidth]{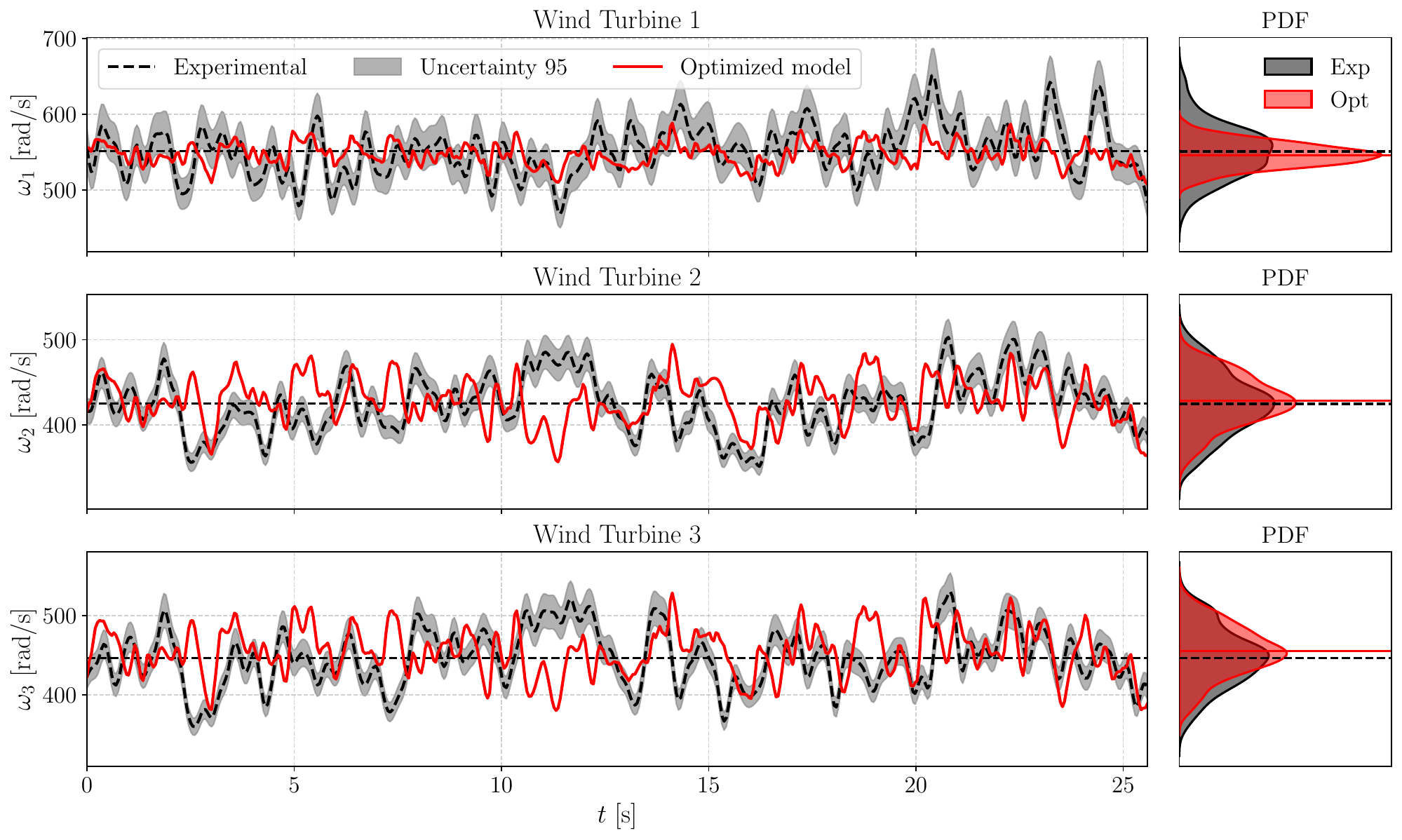}
	\caption{\label{fig:HTresults} Model assimilation results for the three wind turbine models under high-turbulence conditions, using the testing dataset. The continuous red lines represent the adaptive model predictions, the black lines correspond to the experimental data, and the grey shaded bands indicate the associated measurement uncertainty. The plot highlights the substantial fluctuations in rotational speed caused by high turbulence, as well as the inherent limitations of the adaptive model in fully capturing the instantaneous dynamics. Nevertheless, the model accurately reproduces the statistical behavior of the system, particularly the mean and standard deviation of the waked turbines, as shown in the PDF plots on the right.}
\end{figure}

Figure~\ref{fig:CPcomparison2} (top) presents the sampled pairs \( Re - \lambda_1 \) for the first (upstream) turbine and \( \lambda_i - \lambda_{i+1} \) for the downstream (waked) turbines. Samples from the training and test phases are indicated with red and black markers, respectively, while the blue markers represent the states visited by the model when attempting to track the test cases. The orange star indicates the mean value. The dashed line shows the direction of the leading principal component in the sample space. 

Two main observations arise. First, although the model does not perfectly reproduce the instantaneous values, the predictions remain well within the distribution of the observed data, with the mean value largely preserved. Second, the correlation in the sample space is significantly higher for the waked turbines compared to the upstream one. This suggests that, in terms of input/output behavior, the wind farm acts as a low-pass filter on the incoming turbulence.

Figure~\ref{fig:CPcomparison2} (bottom) shows the predicted power coefficient along the principal component of the sampled space, comparing the initial guess and the optimized solutions for both the high-turbulence case and a previous low-turbulence scenario. The observed discrepancy in performance across the three turbines is expected, as wake effects lead to notably lower power coefficients in the downstream turbines. The substantial difference between the initial and optimized predictions in the high-turbulence case highlights the effectiveness of the optimization process. Of particular interest is the sensitivity of the upstream turbine’s power curve to turbulence intensity: high-frequency fluctuations are filtered out by the turbine dynamics, and thus a model trained in low-turbulence conditions---mainly influenced by large-scale fluctuations---tends to yield overly optimistic and potentially nonphysical predictions. Conversely, for the waked turbines, turbulence enhances wake recovery, although only slightly under these conditions, as the increase in the power coefficient relative to the initial guess in the low-turbulence scenario is modest. Consequently, models trained under high-turbulence conditions yield more accurate predictions than those trained in calm flows.

Regarding the shape of the identified power coefficient curves, it is important to consider the sample distribution used for model identification. For example, the apparent double-peaked shape in the upstream turbine’s power curve is likely nonphysical and results from insufficient data in the regions \( \lambda_1 < 4 \) and \( \lambda_1 > 6 \). However, since these operating conditions are rarely visited, this has little impact on the model's predictive performance. Overall, the strong sensitivity of the identified performance curves to the turbulence conditions in the training data underscores the value of continuously adapting the model over the turbine’s operational life. This aligns well with the proposed episodic system identification framework, where the model is not trained once for all but is periodically updated in episodes, each time incorporating newly observed operational data to improve accuracy and robustness.

Finally, Figure \ref{fig:HTresults} shows the model predictions and measured rotational speeds for all three turbines in the test dataset. As in previous figures, adaptive model predictions are plotted in continuous red, experimental measurements in black, and the associated uncertainty as a gray confidence band. The high turbulence environment results in substantial fluctuations in rotational speed across all turbines. For each case, the right panel presents the probability density function (PDF) of the measured data and that predicted by the optimized model. In all cases, the mean is accurately captured, while the standard deviation is slightly underestimated for the upstream turbine, likely due to weaker input–output correlation in the training data (see Figure \ref{fig:CPcomparison2}, top left). Interestingly, the model reconstructs the distributions more accurately for the waked turbines.

\subsection{Control performance from identified models}\label{sec6p3}

The identified model is integrated in the $K\omega^2$ control law as described in \ref{sec:scenario3} for both the upstream and downstream turbine in the low turbulence scenario.
The control challenge consist in keeping a reference trajectory in terms of tip-speed ratio, hence define a time varying reference $\lambda$ (and associated power coefficient) in \eqref{K_eq} rather than the power maximization. 

\begin{figure}[H]
	\centering
	\includegraphics[width=\textwidth]{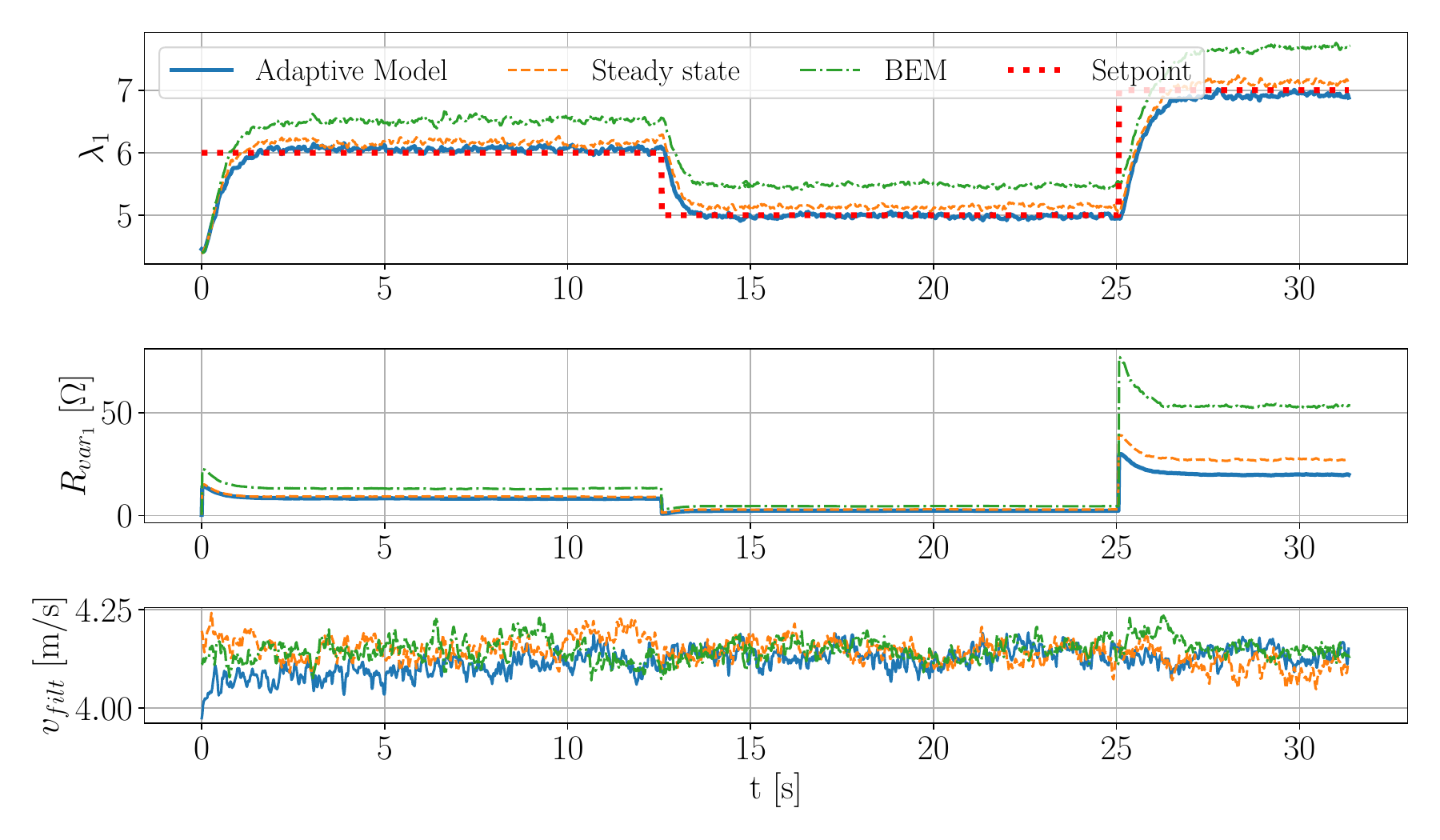}
	\caption{\label{fig:control1} Test of the adaptive model of the upstream turbine in the low turbulence scenario using the $K\omega^2$ control law. The setpoint in terms of tip speed ratio is highlighted in red. The controller using the adaptive model (blue) is compared with the controller using the model obtained through BEM simulations (green) and steady-state data regression used as an initial guess for the optimization (orange). It can be seen that the controller with the adaptive model outperforms the others.}
\end{figure}

The results for the upstream turbine are shown in Figure~\ref{fig:control1}. The top plot displays the dynamic set point alongside the actual tip-speed ratio for three controllers based on different models: (1) the Blade Element Momentum (BEM)-regressed model (green), (2) the initial guess power coefficient curve assuming steady-state conditions (orange), and (3) the adapted model obtained after the identification process (blue). The middle plot shows the control action, i.e., the variation in generator resistance applied by each controller, while the bottom plot presents the incoming wind velocity during the three control tests, which remained relatively similar. It is worth noting that the large variations in set point were not intended to mimic a realistic control task—which would typically aim to track the optimal power coefficient—but were instead designed to stress-test the controllers.

The superior performance of the controller driven by the adapted model is evident. The large steady-state errors and excessive actuation observed in the BEM-driven controller can be attributed to model deficiencies, stemming from the uncertain prediction of low-Reynolds-number effects. These uncertainties affect the airfoil performance estimates at small testing scales, which are then used as inputs to the BEM to compute the power coefficient. Although these could be corrected by more sophisticated control laws, it is interesting to note that the adaptive formulation yields excellent tracking performances with the simplest $K\omega^2$ controller. 

The difference in control performances of the initial and adapted model appears minor in comparison. After all, as described in the previous section, the initial guess yields overall satisfactory models as long as the stability properties are preserved. Nevertheless, the difference is appreciable. Considering the impact of power curve modeling on energy production \citep{pao2009tutorial}, and considering that wind turbine dynamics change over their lifespan due to factors such as mechanical wear, environmental conditions, and blade fouling, the ability to adapt and refine the model remains essential for maintaining optimal performance and long-term efficiency.

\begin{figure}[H]
	\centering
	\includegraphics[width=\textwidth]{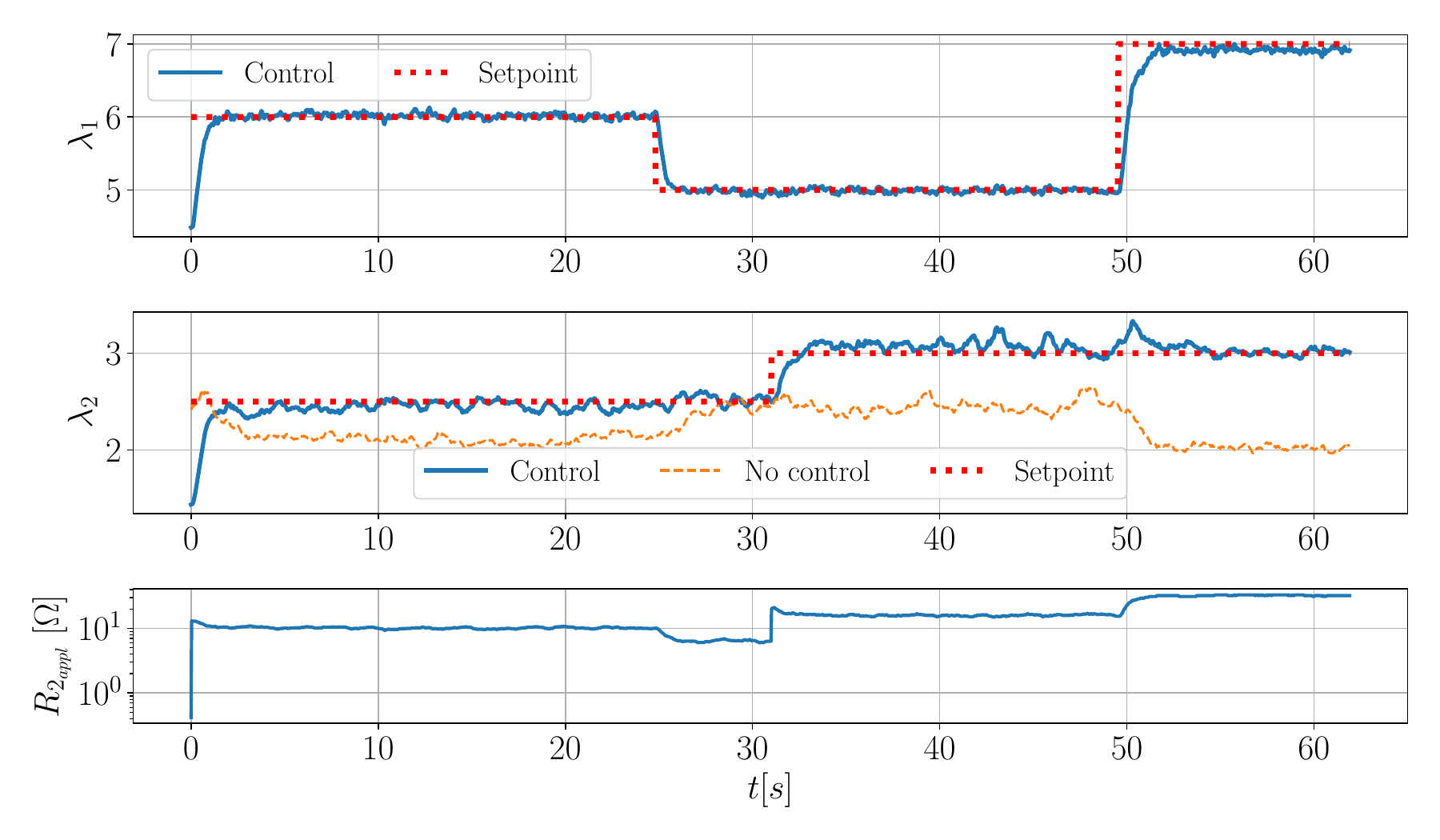}
	\caption{\label{fig:control2} Test of the adaptive models of the tandem configuration in the low turbulence scenario with the $K\omega^2$ control law. Both the first wind turbine and the downstream turbine are controlled to follow a determined setpoint highlighted in red. A comparison with the uncontrolled case of the downstream turbine (orange) is presented. In this scenario, the effect of the wake is clearly visible, as is the corrective action applied by the controller to the downstream turbine to counteract this effect.}
\end{figure}

Finally, a similar analysis is presented for the waked turbine in Figure~\ref{fig:control2}, a scenario well beyond the reach of traditional modeling approaches due to strong inflow unsteadiness and limited flow characterization. Time-dependent set points are prescribed for both turbines. The top plot shows the tracking performance of the upstream turbine, while the second plot presents the response of the downstream turbine, alongside an uncontrolled baseline scenario. The uncontrolled response clearly reveals the wake effect on the second turbine when the upstream set point $\lambda_1$ is changed at $t=25$ s and $t=50$ s. The third plot shows the control input, with a visible actuation step triggered by the set point change, as well as the downstream response to upstream variation.

The key result is highlighted in the second plot: the controller, based on the identified model, successfully mitigates the impact of upstream disturbances (at $t=25$ s and $t=50$ s) and accurately tracks the reference step change (at $t=32$ s). Although some oscillations appear in the controlled $\lambda_2$ signal, the overall results demonstrate excellent tracking and disturbance rejection performance within the model bandwidth, even in the presence of wake-induced disturbances. These oscillations can be attributed to an inherent limitation of the $K\omega^2$ controller, which, by design, assumes steady-state operating conditions (see \citet{pao2009tutorial}) and lacks feedback control command from the incoming wind velocity. Consequently, its tracking performances degrade under rapidly changing wind conditions because changes in wind velocity are only sensed via changes in rotor speed, which are lagged due to inertia. This limitation is visible for the wakes turbine but not for the upstream one, which operates in low turbulence and slow variations of the wind speed.

\section{Conclusions} \label{sec:6}
In this work, we proposed and experimentally validated a system identification approach for free-stream and waked wind turbines based on the real-time estimation of nonlinear power coefficient curves. By combining compactly supported radial basis functions with low-order polynomial regressor, the method achieves accurate and physically interpretable representations of turbine performance under dynamic conditions. The methodology was tested on both low and high turbulence scenarios using small-scale wind turbines, with the identified models showing strong agreement with experimental measurements.

The adaptive models were successfully integrated into the $K\omega^2$ controller, demonstrating enhanced performance over conventional control strategies based on BEM or steady-state regressions. In a controlled tandem configuration, the adaptive controllers were able to mitigate wake-induced effects on a downstream turbine, validating the potential of this framework for coordinated wind farm operation. Moreover, the ability of the model to adapt to different operating regimes, including noisy and turbulent conditions, highlights its robustness and practical applicability. The proposed identification process, based on adjoint gradient computation and ADAM optimization, enables real-time adaptability while remaining computationally efficient. Importantly, the models can be directly embedded into already existing model-based industrial controllers without requiring architectural changes.

Future developments could explore the integration of this adaptive modeling approach into more advanced control strategies such as Model Predictive Control (MPC), as well as its extension to full-scale wind farms with lidar-based wind measurements. Overall, this study illustrates a practical, data-driven pathway toward improved wind turbine performance through interpretable and flexible model identification.

\section*{Acknowledgments}
This research was initially developed as part of the research master’s project of Sebastiano Randino, supported by a VKI NATO Fellowship. It was subsequently refined and extended during the PhD work of Sebastiano Randino, funded by a FNRS FRIA Fellowship (Fellowship number 40029825). Lorenzo Schena is funded by an FWO Fellowship (number 1567925N). The authors gratefully acknowledge the contribution of Jacquet Charles, who worked on the farm testing in the framework of his research master project. This work is part of the RE-TWIST project, which has received funding from the European Research Council (ERC) under the European Union’s Horizon Europe programme (grant agreement No 101165479). The views expressed are those of the authors and do not necessarily reflect those of the European Union or the ERC.



\appendix


\clearpage		

\end{document}